\documentclass[twocolumn,showpacs,pra,superscriptaddress,longbibliography]{revtex4-1}
\usepackage{times}
\usepackage{amsmath}
\usepackage{amssymb}
\usepackage{graphicx}
\usepackage[unicode]{hyperref}
\hypersetup{
   unicode=true,          % non-Latin characters in AcrobatÕs bookmarks
   plainpages=false,
   colorlinks=true,       % false: boxed links; true: colored links
   citecolor=blue,        % color of links to bibliography
}
\usepackage[caption=false,position=top,singlelinecheck=off,justification=raggedright]{subfig}
\begin{document}

\title{Hierarchical relaxation dynamics in a tilted two-band Bose-Hubbard model}

\author{Jayson G. Cosme}

\affiliation{Dodd-Walls Centre for Photonics and Quantum Technology, New Zealand Institute for Advanced Study, Centre for Theoretical Chemistry and Physics, Massey University Auckland, Private Bag 102904, North Shore, Auckland 0745, New Zealand}
\affiliation{Zentrum f\"{u}r Optische Quantentechnologien and Institut f\"{u}r Laserphysik, Universit\"{a}t Hamburg, 22761 Hamburg, Germany}
\affiliation{The Hamburg Centre for Ultrafast Imaging, Luruper Chaussee 149, 22761 Hamburg, Germany}


\date{\today}
\begin{abstract}
We numerically examine slow and hierarchical relaxation dynamics of interacting bosons described by a tilted two-band Bose-Hubbard model. The system is found to exhibit signatures of quantum chaos within the spectrum and the validity of the eigenstate thermalization hypothesis for relevant physical observables is demonstrated for certain parameter regimes. Using the truncated Wigner representation in the semiclassical limit of the system, dynamics of relevant observables reveal hierarchical relaxation and the appearance of prethermalized states is studied from the perspective of statistics of the underlying mean-field trajectories. The observed prethermalization scenario can be attributed to different stages
of glassy dynamics in the mode-time configuration space due to dynamical phase transition between ergodic and nonergodic trajectories. 
\end{abstract}
\maketitle

\section{Introduction }

Recent developments and experimental efforts in the field of ultracold atomic physics \cite{Bloch2008,Kinoshita2006,Gring2012,Langen2015} have paved the way for explorations of nonequilibrium dynamics in isolated quantum systems \cite{Polkovnikov2003}. In closed quantum systems, thermalization can be understood as equilibration of local or physical observables to a value described by a thermal ensemble \cite{Srednicki1994,Deutsch1991,Rigol2008}. The eigenstate thermalization hypothesis (ETH), states that the long-time average of a local observable will tend to a microcanonical ensemble prediction \cite{Srednicki1994,Deutsch1991,Rigol2008}. The issue of thermalization under unitary evolution is closely related to the question of integrability of the Hamiltonian that describes the quantum system. In particular, the onset of statistical relaxation is shown to coincide with the breakdown of integrability indicated by the onset of quantum chaos in the spectrum of the final Hamiltonian \cite{Alessio2016,Santos2012,Santos2012c}.
However, an important consideration in the dynamical process of relaxation is the possibility of multistage dynamics due to the appearance of intermediate nonthermal stationary states. Indeed, prethermalization dynamics of various systems have been observed in experimental \cite{Gring2012,Smith2013,Langen2015}
and theoretical  \cite{Moeckel2008,Kollar2011,Marcuzzi2013,Essler2014,Nessi2014,Landea2015,Bertini2015,Nessi2015,Cosme2015,Marcuzzi2016,Bertini2016,Chiocchetta2017,Alba2017}
works alike. A recent review on this topic of prethermalization specially for quantum systems which are nearly inegrable can be found in Ref.~\cite{Langen2016}.

Phase space methods have been successfully utilized to study various non-equilibrium dynamical phenomena such as superfluid flow \cite{Mathey2014}, many-body spin dynamics \cite{Schachenmayer2015}, and breather relaxation \cite{Opanchuk2017} just to name a few. More importantly for the purpose of this work, it has been used to provide new insights in relaxation \cite{Chianca2011,Mathew2017}, prethermalization \cite{Landea2015}, and thermalization dynamics \cite{Larson2013,Altland2012prl,Altland2012njp,Cosme2014,Acevedo2017} in closed quantum systems.
In Refs. \cite{Altland2012prl,Altland2012njp}, it was conjectured that thermalization is manifested in uniformly distributed quasiprobability distributions as they spread over the available energy shell. Furthermore, it was numerically demonstrated that the marginal distributions of a phase-space distribution, the Wigner function, are consistent with the corresponding microcanonical distributions \cite{Cosme2014}. More recently in Ref.~\cite{Landea2015}, features of the ensemble of classical trajectories are found to share some resemblance with properties of the ensemble of histories in the relaxation of glassy systems. The similarities include the existence of a dynamical phase transition between ergodic and nonergodic phases and also the occurrence of dynamical heterogeneity in the space-time structure of individual trajectories \cite{Garrahan2007,Hedges2009,Garrahan2010,Chandler2010}.

In this work, we present a case study of quantum relaxation in a tilted Bose-Hubbard model through phase space representation of the quantum dynamics. Specifically,
we adopt a two-band Bose-Hubbard model studied in \cite{Tomadin2008,Plotz2010,Plotz2011,Parra-Murillo2013,Parra-Murillo2014} and demonstrate quantum equilibration within this model. The interband coupling adds another complexity in the single band tilted system, which already possesses rich quantum and mean-field dynamics including, for example, the emergence of nonlinear chaos \cite{Thommen2003,Vermersch2015}, wave packet spreading \cite{Krimer2009}, and the presence of both stable and unstable Bloch oscillations \cite{Kolovsky2009,Kolovsky2010}. In the two-band model, we investigate the manifestation of quantum chaos in the energy spectra and the validity of the ETH is investigated for small system sizes. Then, for large number of bosons, we show that the system can exhibit both fast single-step or slow multi-step relaxation process of physical observables. The observed multi-step relaxation motivates us to characterize the statistics of the mean-field trajectories which are used to capture the initial quantum noise of the system. Moreover, we provide numerical evidence that the hierarchical relaxation of single particle observables is correlated with the coexistence of ergodic and nonergodic phases in the ensemble of classical trajectories. 

This paper is organized as follows. 
In Sec.~\ref{sec:system}, we introduce the model used to describe the dynamics of a tilted optical lattice with the possibility of occupying the second energy band of the lattice. We also show the equations of motion used in calculating the expectation value of operators in the truncated Wigner formalism. The appearance of quantum chaos within the spectrum of the system is explored in Sec.~\ref{sec:levelspace}. In Sec.~\ref{sec:eth}, we investigate the validity of ETH for relevant observables. In Sec.~\ref{sec:equil}, we present numerical results on the quantum equilibration of a one-body local observable. We further investigate the relationship between the relaxation process and properties of the underlying mean-field trajectories in Sec.~\ref{sec:mfield}. Lastly, we summarize the findings of this work in Sec.~\ref{sec:conc}.

\section{System and Equations of Motion}\label{sec:system}

\subsection{Model}

We consider a two-band Bose-Hubbard model with additional external Stark force as in Refs.~\cite{Tomadin2008,Plotz2010,Plotz2011}.
A similar model but for an array of double wells constructed by means of superlattices was recently studied in Refs.~\cite{Parra-Murillo2013,Parra-Murillo2014}.
For simplicity, we study the ubiquitous single lattice model described by the single-particle Hamiltonian
\begin{equation}\label{eq:hsp}
 \hat{H}_{\mathrm{sp}} = \frac{-\hbar^2}{2m}\frac{\partial^2}{\partial x^2} + V_0 \mathrm{cos}(2\pi x/d) + Fx. 
\end{equation}
We refer to Ref.~\cite{Tomadin2008} for details on calculating the Wannier functions and the corresponding many-body parameters of the resulting tight-binding Hamiltonian.
In contrast to previous studies, we use open or hard-wall boundary conditions in the tight-binding model given by
\begin{align}\label{eq:Hamilt}
	&\hat{H} = \sum^{2}_{\ell=1}\sum^L_{r=1} \biggl [ (E^\ell + rF ) \hat{n}^{\ell}_{r} + \frac{g}{2}W^{\ell}\hat{n}^{\ell}_{r}(\hat{n}^{\ell}_{r}-1) \biggr] \\ \nonumber
	& - \sum^{2}_{\ell=1}\sum^{L-1}_{r=1} J^{\ell}( \hat{b}^{\ell\dagger}_{r}\hat{b}^{\ell}_{r+1} + \mathrm{H.c.}) + FC^{12}\sum^{L}_{r=1}( \hat{b}^{1\dagger}_{r}\hat{b}^{2}_{r} + \mathrm{H.c.}) \\ \nonumber
	&+\frac{gW^{12}}{2}\sum^{L}_{r=1}\biggl[ 4\hat{n}^{1}_{r}\hat{n}^{2}_{r}  + (\hat{b}^{1\dagger}_{r}\hat{b}^{1\dagger}_{r}\hat{b}^{2}_{r}\hat{b}^{2}_{r} + \hat{b}^{2\dagger}_{r}\hat{b}^{2\dagger}_{r}\hat{b}^{1}_{r}\hat{b}^{1}_{r})\biggr],
\end{align}
where $\hat{b}^{\ell\dagger}_{r}(\hat{b}^{\ell}_{r})$ are bosonic creation (annihilation) operators of an atom in band $\ell \in [1,2]$ and site $r \in [1,L]$. We refer to a pair of $\{\ell,r\}$ as a mode with the total number of modes $\mathcal{M}=2 \times L$. The on-site energies are $(E^\ell + rF )$. The tunneling terms are given by $J^\ell$. $W^{12}$ denotes the pair interband tunneling term. There is also a single particle interband coupling given by $C^{12}$. Finally, the interaction terms between atoms on the same site and the same band are denoted by $W^{\ell}$. The interaction strength $g$ is assumed to be an adjustable parameter. 

For the rest of this work, we measure energy in units of the recoil energy $E_R=\hbar^2\pi^2/(2 m d^2)$. Also, the following quantities are measured accordingly: (i) time $t$ in $T=\hbar/E_R$; (ii) spatial dimension $x$ in terms of the lattice spacing $d$; (iii) interaction $g$ in $E_Rd$; and (iv) $F$ in $E_R/d$.
We restrict our calculations using a set of parameters for the optical lattice potential in Eq.~\eqref{eq:hsp}. Specifically, we set $d=1$, $V_0=2$ and $F=(E^2-E^1)/5$. Note that this chosen tilt($F=0.5932$) is close to the resonance condition between the upper band of the first site and the lower band of the fifth site $r=5$ ($F=0.5956$) \cite{Plotz2010}. The resulting Hamiltonian coefficients are: $(E^2-E^1)=2.966$, $C^{12}=-0.183$, $J^1=0.086$, $J^2=-0.462$, $W^1/W^2 = 1.665$, and  $W^1/W^{12} = 2.381$. When fixing the trap parameters, the only remaining adjustable parameter is the interaction strength $g$. In the subsequent simulations, we shall vary $g$ in order to probe the onset of quantum chaos and the emergence of nonlinear behavior in the mean-field trajectories which can lead to effective relaxation in the system.

\subsection{Equations of Motion}

One of the main ideas of phase space methods is to represent quantum operators as complex numbers spanning the phase space of canonical variables $\psi$ and $\psi^*$. In this work, we shall focus on one representation in accordance with the Wigner-Weyl quantization approach \cite{Polkovnikov2010}. This can be achieved by mapping the von Neumann equation for the density matrix $\hat{\rho}(t)$, 
\begin{equation}\label{eq:vNe}
	i\hbar\frac{\partial \hat{\rho}(t)}{\partial t} = [\hat{H},\hat{\rho}(t)],
\end{equation}
to a corresponding evolution equation for the Wigner function $W(t)$, which is just the Wigner-Weyl transform of the density matrix \cite{Polkovnikov2010}. The Wigner function is a quasiprobability distribution in the phase space containing the canonical variables. By taking the Wigner-Weyl transform of both sides of Eq.~\eqref{eq:vNe} in the coherent state representation \cite{Polkovnikov2010}, one can obtain  
\begin{equation}\label{eq:fwa}
	i\hbar \frac{\partial W}{\partial t} =  \{H_W,W\}_{\mathrm{MBC}},
\end{equation}  
where $H_W$ is the Weyl symbol of the Hamiltonian operator and $\{\cdots\}_{\mathrm{MBC}}$ denotes the Moyal bracket \cite{Polkovnikov2010}. 
One useful trick in evaluating the Weyl symbol or equivalently the Wigner-Weyl transformation of the Hamiltonian is by using the coherent state Bopp representation \cite{Polkovnikov2010}
\begin{align}
\hat{b}^{\ell\dagger}_{r} &\to {b}^{\ell*}_{r}-\frac{1}{2}\frac{\partial}{\partial {b}^{\ell}_{r}} \\ \nonumber
\hat{b}^{\ell}_{r} &\to {b}^{\ell}_{r}+\frac{1}{2}\frac{\partial}{\partial {b}^{\ell*}_{r}}.
\end{align}
Then, the Weyl symbol of the Hamiltonian $H_W$ can be simply obtained by substituting the above representation to the bosonic operators in the tight-binding Hamiltonian in Eq.~\eqref{eq:Hamilt}. The Moyal bracket is defined as \cite{Polkovnikov2010}
\begin{equation}
\{H_W,W\}_{\mathrm{MBC}}=H_W\mathrm{exp}\left(\frac{\Lambda_C}{2}\right)W,
\end{equation} 
where the Poisson bracket operator is 
\begin{equation}
H_W\Lambda_C W = \{H_W,W\}_{\mathrm{C}}=\sum_{\ell,r} \frac{\partial H_W}{\partial {b}^{\ell}_{r}}\frac{\partial W}{\partial {b}^{\ell*}_{r}}-\frac{\partial H_W}{\partial {b}^{\ell*}_{r}}\frac{\partial W}{\partial {b}^{\ell}_{r}}.
\end{equation} 
The Moyal bracket can be expanded in $1/N$ up to the leading order in the limit of large number of bosons $N$. This truncation is equivalent to neglecting the third-order derivatives $\frac{\partial^3H_W}{\partial \psi_\alpha\partial \psi_\beta\partial \psi_\gamma}$. This simplification is called the truncated Wigner approximation (TWA) \cite{Blakie2008,Polkovnikov2010} and the resulting evolution equation reminiscent of the classical Liouville equation is
\begin{equation}\label{eq:twa}
		i\hbar \frac{\partial W}{\partial t} =  \{H_W,W\}_{\mathrm{C}}.
\end{equation}
The equations of motion are obtained by noting that the Wigner function is conserved along the classical trajectories satisfying
\begin{equation}\label{eq:seom}
	i\hbar \frac{\partial \psi_j}{\partial t} = \frac{\partial H_W}{\partial \psi^*_j}.
\end{equation} 
In this formalism, the expectation value of an operator at a specific time can be calculated as,
\begin{equation}
 \langle \hat{\Omega}(t) \rangle = \int d\psi_0 d\psi^*_0 W(\psi_0,\psi^*_0)\Omega_W(\psi(t),\psi^*(t),t),
\end{equation}
where $W(\psi_0,\psi^*_0)$ is the initial Wigner function and $d\psi_0d\psi^*_0 = \prod_j d\psi_j(0)d\psi^*_j(0)$.
In practice, $\langle \hat{\Omega}(t) \rangle$ is obtained by solving the set of coupled differential equations in Eq.~\eqref{eq:seom} using an ensemble of initial conditions, which accurately samples the quantum noise of a chosen initial state, then averaging afterwards. 

We map the bosonic creation and annihilation operators to the corresponding complex $c$-numbers. The correspondence between the quantum operators and the complex $c$-numbers can be easily calculated using the Bopp representation $\hat{b}^{ \ell\dagger}_{r} \to b^{\ell*}_{r} - \frac{1}{2}\frac{\partial}{\partial b^{\ell}_{r}}$ and $\hat{b}^{ \ell}_{r} \to b^{\ell}_{r} + \frac{1}{2}\frac{\partial}{\partial b^{\ell*}_{r}}$.
The equations of motion for the $c$-numbers of Eq.~\eqref{eq:Hamilt} reads
\begin{align}\label{eq:eom}
	&i\hbar \frac{\partial b^{\ell}_{r}}{\partial t} = (E^{\ell}+rF)b^{\ell}_{r} + FC^{12}b^{\ell'}_{r} \\ \nonumber
	&- J^{\ell}\biggl( b^{\ell}_{r+1}(1-\delta_{r,L})+b^{\ell}_{r-1}(1-\delta_{r,1}) \biggr) + 2gW^{12}b^{\ell}_{r}|b^{\ell'}_{r}|^2 \\ \nonumber
	&+ gW^{12}b^{\ell*}_{r}(b^{\ell'}_{r})^2 +  \frac{gW^{\ell}}{2}\biggl(2|b^{\ell}_{r}|^2 b^{\ell}_{r} -b^{\ell}_{r}\biggr), 
\end{align}
where $\ell \neq \ell'$ and $\delta_{i,j}$ is the Kronecker delta, which arises from imposing open boundary condition. Note that in the single-band limit of the model, Eq.~\eqref{eq:eom} simplifies to the discrete nonlinear Sch\"{o}dinger equation found in Ref.~\cite{Krimer2009}. The set of coupled differential equation in Eq.~\eqref{eq:eom} is numerically integrated using standard Runge-Kutta algorithm. 

The ensuing dynamics will also depend on the particular choice of initial state or equivalently the initial Wigner distribution. We are interested in the dynamics of two kinds of initial states: (i) coherent state and (ii) Fock state.
We use the appropriate sampling of the initial Wigner function when solving the coupled differential equations Eq.~\eqref{eq:eom} \cite{Blakie2008,Olsen2009}.
For a coherent state corresponding to a condensed state of bosons, the ensemble of initial conditions is sampled as $b^{\ell}_r=\sqrt{N^{\ell}_r(t=0)}+\frac{1}{2}(\nu_1+i \nu_2)$ where $\nu_i$ are Gaussian random variables distributed with $\overline{\nu_i}=0$ and $\overline{\nu_j \nu_k}=\delta_{j,k}$. 
Modes that are initially unoccupied or the so-called vacuum state will also follow this sampling procedure.
Fock states of definite occupation number $N^{\ell}_r$, such that $\sum N^{\ell}_r = N$, are sampled as $b^{\ell}_r=(p+q\nu)e^{i2\pi\xi}$ 
where $\nu$ is a Gaussian random variable and $\xi$ is a uniform random variable in the interval $[0,1]$, 
$p=\frac{1}{2}(2N^{\ell}_r+1+2\sqrt{(N^{\ell}_r)^2+N^{\ell}_r})^{1/2}$ and $q=1/(4p)$. 
This approximation is valid for large occupation number in a mode. Moreover, this sampling correctly captures the mean and the variance of the exact Wigner function and it can be shown to generate the correct moments up to a correction of order $1/N^2$ \cite{Olsen2009}.

%\section{Choice of Parameters and Initial States}\label{sec:levelspaceth}

\section{Emergence of quantum chaos in the spectrum}\label{sec:levelspace}

Here, we elucidate the choice of parameters, i.e., the number of lattice sites $L$ and the rescaled interaction strength $Ng$, for the simulations of quantum dynamics.  To this end, we rely on the relation between the onset of statistical relaxation and quantum chaos \cite{Alessio2016,Santos2012,Santos2012c}.
We characterize the presence of quantum chaos using the distribution of the ratio of consecutive gaps \cite{Oganesyan2007,Atas2013}.
For a time-independent Hamiltonian, the chaotic regime is conjectured to possess similar level statistics as the Gaussian orthogonal ensemble (GOE), which is the Wigner-Dyson distribution \cite{Brody1981,Montambaux1993}. On the other hand, integrable models are expected to exhibit Poissonian level statistics \cite{Brody1981,Berry1977}.

\begin{figure}[!ht]
\begin{center}
\includegraphics[width=1\columnwidth]{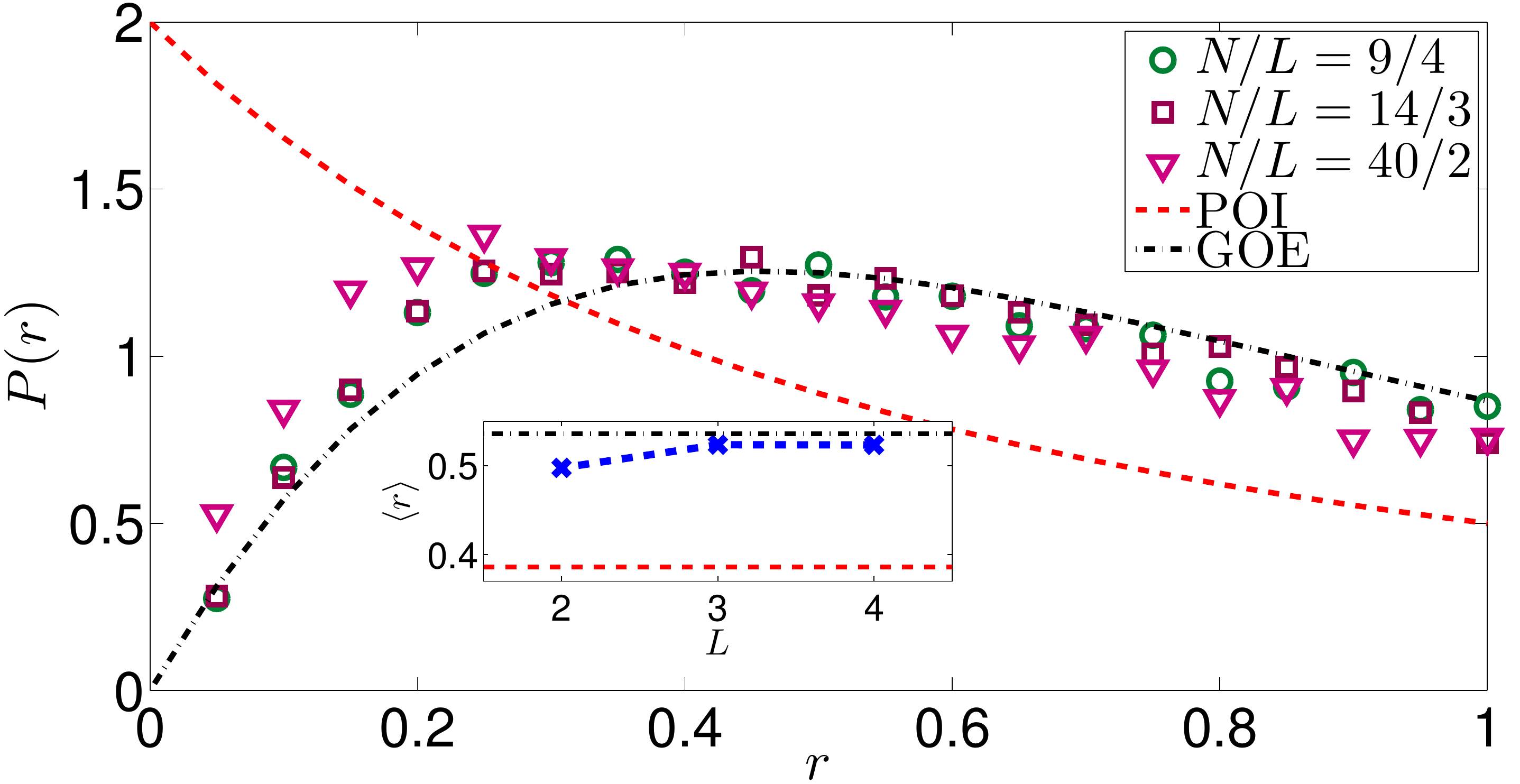}
\includegraphics[width=1\columnwidth]{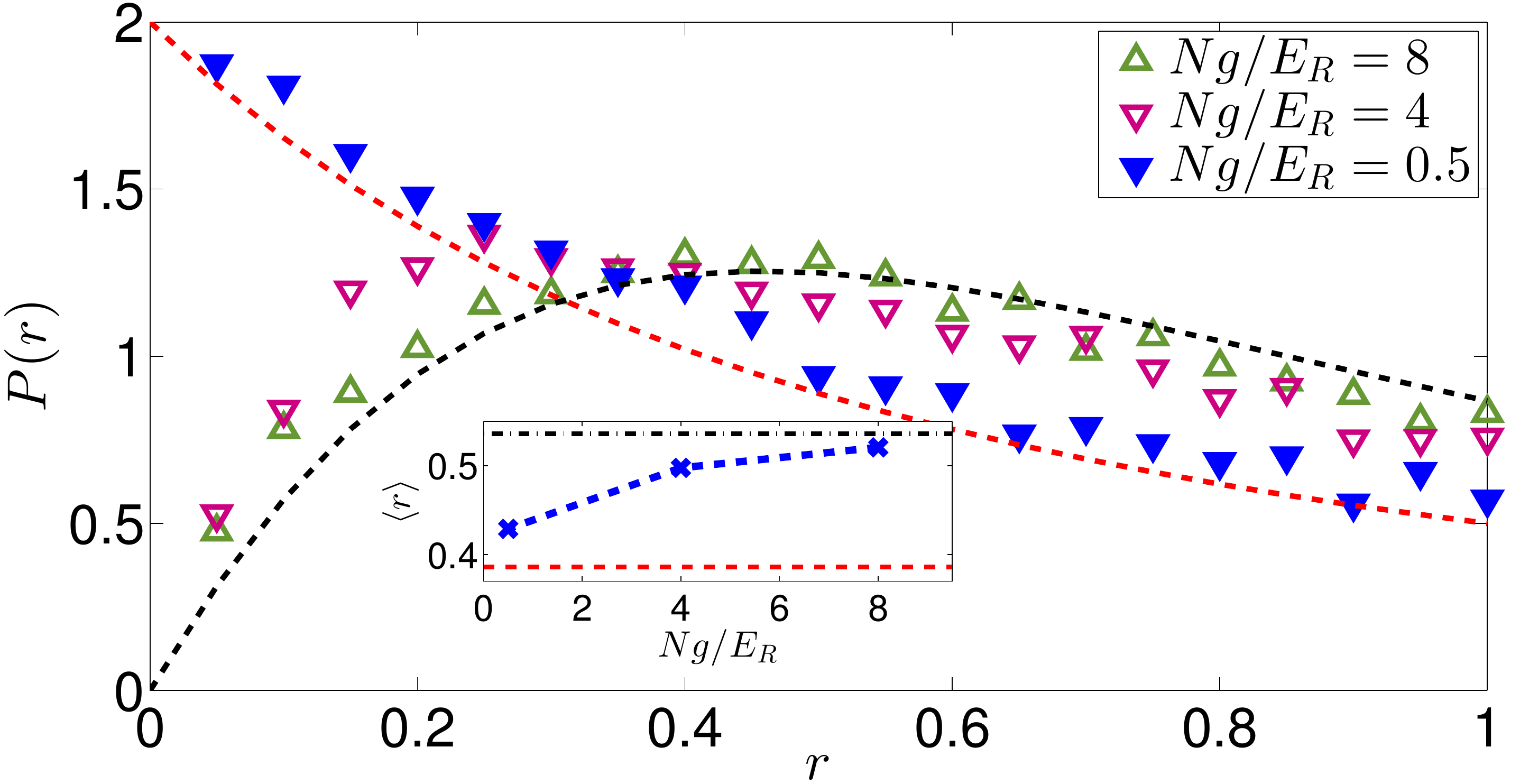}
\protect\caption{(Color online) Distribution of the ratio of adjacent level spacings $P(r)$. (Top) Different system size $L$ and fixed interaction $Ng/E_R=4$; (Bottom) For $N=40$, $L=2$, and varying interaction strengths.
The GOE (dark-dashed) and the Poissonian (bright-dashed) averages are shown for comparison. (Inset) Mean value $\langle r \rangle$ as a function of: (Top) $L$ and; (Bottom) $Ng/E_R$ in the lower panel.}
\label{fig:lsr}
\end{center}
\end{figure}

Due to the possibility of probing large interaction strengths,  i.e., $Ng/E_R J^{\ell} > 1$, it is convenient to avoid any unfolding procedure of the spectrum \cite{Kollath2010}. Consider the set of level spacings $\{s_k\} = \{E_{k+1}-E_{k}\}$ where $\{E_k\}$ is the set of eigenenergies in ascending order.
We obtain the distribution of the ratio of consecutive gaps between adjacent levels, 
$r_k = \frac{\mathrm{min}(s_k,s_{k-1})}{\mathrm{max}(s_k,s_{k-1})} = \mathrm{min}(\frac{s_k}{s_{k-1}},\frac{s_{k-1}}{s_{k}})$ \cite{Oganesyan2007, Kollath2010, Atas2013}.
Firstly, we obtain the full spectrum of the Hamiltonian using exact diagonalization. We set the number of bosons depending on the number of sites such that the dimension of the Hilbert space is around $D \sim 10^4$ to allow for better spectral statistics. In particular, we choose $N=40$ for $L=2$, $N=14$ for $L=3$, and $N=9$ for $L=4$.
We compare the distribution $P(r)$ obtained from exact diagonalization with the analytical expressions for the Poisson and the GOE distributions \cite{Atas2013},
\begin{equation}
P_\mathrm{P}(r)=\frac{2}{(1+r)^2},~P_\mathrm{GOE}(r)=\frac{27}{4}\frac{r+r^2}{(1+r+r^2)^{5/2}}.
\end{equation}
A more quantitative comparison can be made from the mean value $\langle r \rangle$,
which is $\langle r \rangle_\mathrm{P} = 2\mathrm{ln}2 - 1 \approx 0.3863$ for the Poisson distribution and
$\langle r \rangle_\mathrm{GOE} \approx 0.5359$ for the GOE \cite{Atas2013}.

In Fig.~\ref{fig:lsr}, we compare $P(r)$ to the Poisson and GOE distributions. It can be seen that $P(r)$ approaches the GOE prediction already for $L=3$. Moreover, for fixed number of bosons and lattice size, i.e., $N=40$ and $L=2$, the agreement of $P(r)$ with the GOE prediction improves when the interaction strength $g$ is increased as shown in Fig.~\ref{fig:lsr}. Since our main interest lies on the relaxation dynamics of the system, we focus on the strongly interacting regime, i.e., $Ng/E_R=4$ and $Ng/E_R=8$. On the one hand, the case $Ng/E_R=4$ represents the scenario when the Hamiltonian is not yet fully chaotic albeit level repulsion is already present. On the other hand, the case $Ng/E_R=8$ possesses a fully chaotic spectrum with $P(r)$ closely resembling the GOE prediction. It is noteworthy to mention that nonintegrability of a system is correlated with emergence of quantum chaos in the spectrum of the system. Thus, in the fully quantum chaotic regime of $Ng/E_R=8$, relevant observables in the system are expected to thermalize according to the ETH. Although recently, there are some observables that are shown to violate the ETH even if the underlying Hamiltonian is nonintegrable \cite{Garrison2017}

\section{Eigenstate thermalization hypothesis}\label{sec:eth}

According to ideas of the eigenstate thermalization hypothesis (ETH), the long-time average of a local observable $\hat{A}$ is equivalent to a corresponding microcanonical ensemble prediction if the eigenstate expectation values (EEV), $\langle \hat{A} \rangle _{kk}=\langle k | \hat{A} |k \rangle$, are narrowly distributed in the eigenenergies $E_k$ of the system \cite{Deutsch1991,Srednicki1994,Rigol2008}. Various theoretical studies have demonstrated that validity of the ETH coincides with quantum ergodicity in closed quanutm systems (see Ref.~\cite{Alessio2016} and references therein). For our purpose, the distribution of EEV will serve as a guide on the optimal choice of initial energy and consequently initial states for which the system might exhibit relaxation of local observables. In particular, we are interested in two classes of observables namely the mode-occupation number $\langle \hat{n}^{\ell}_{r} \rangle$ and the upper-band occupation number $\langle M \rangle=\langle \sum_r \hat{n}^{2}_r \rangle$, which is also known as manifold \cite{Parra-Murillo2013,Parra-Murillo2014}.

\begin{figure}[!ht]
\begin{center}
\includegraphics[width=0.5\columnwidth]{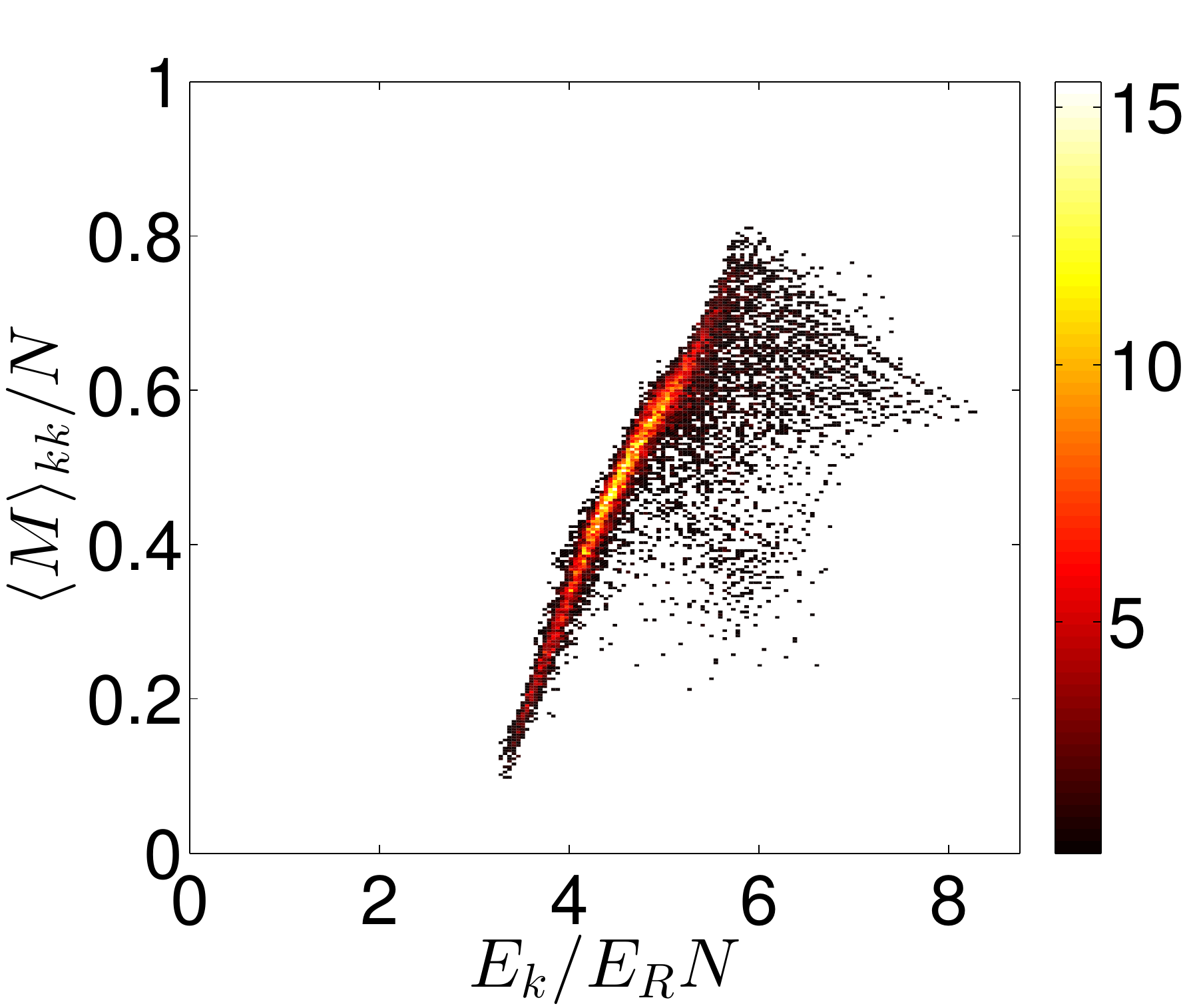}\includegraphics[width=0.5\columnwidth]{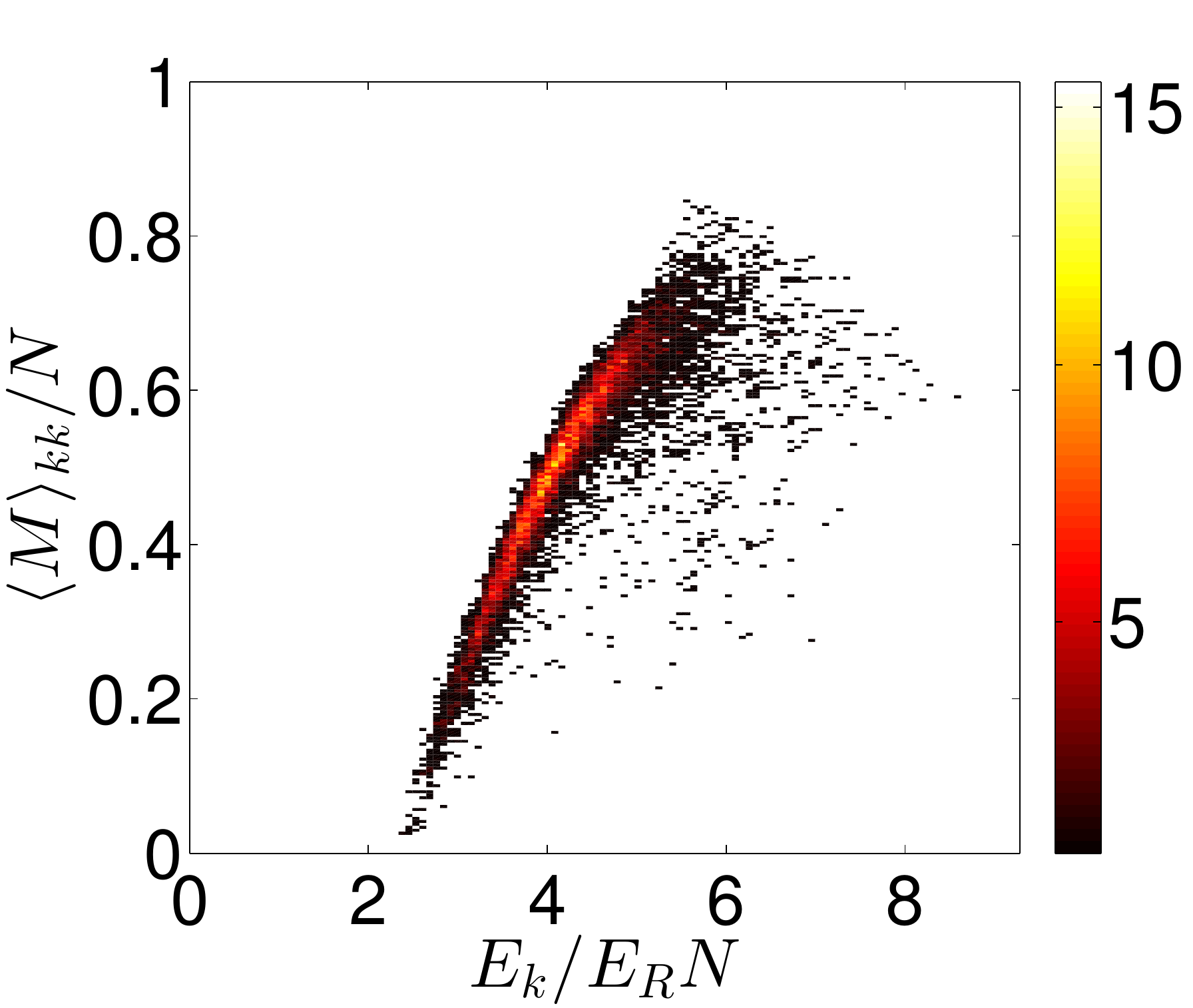}
\includegraphics[width=0.5\columnwidth]{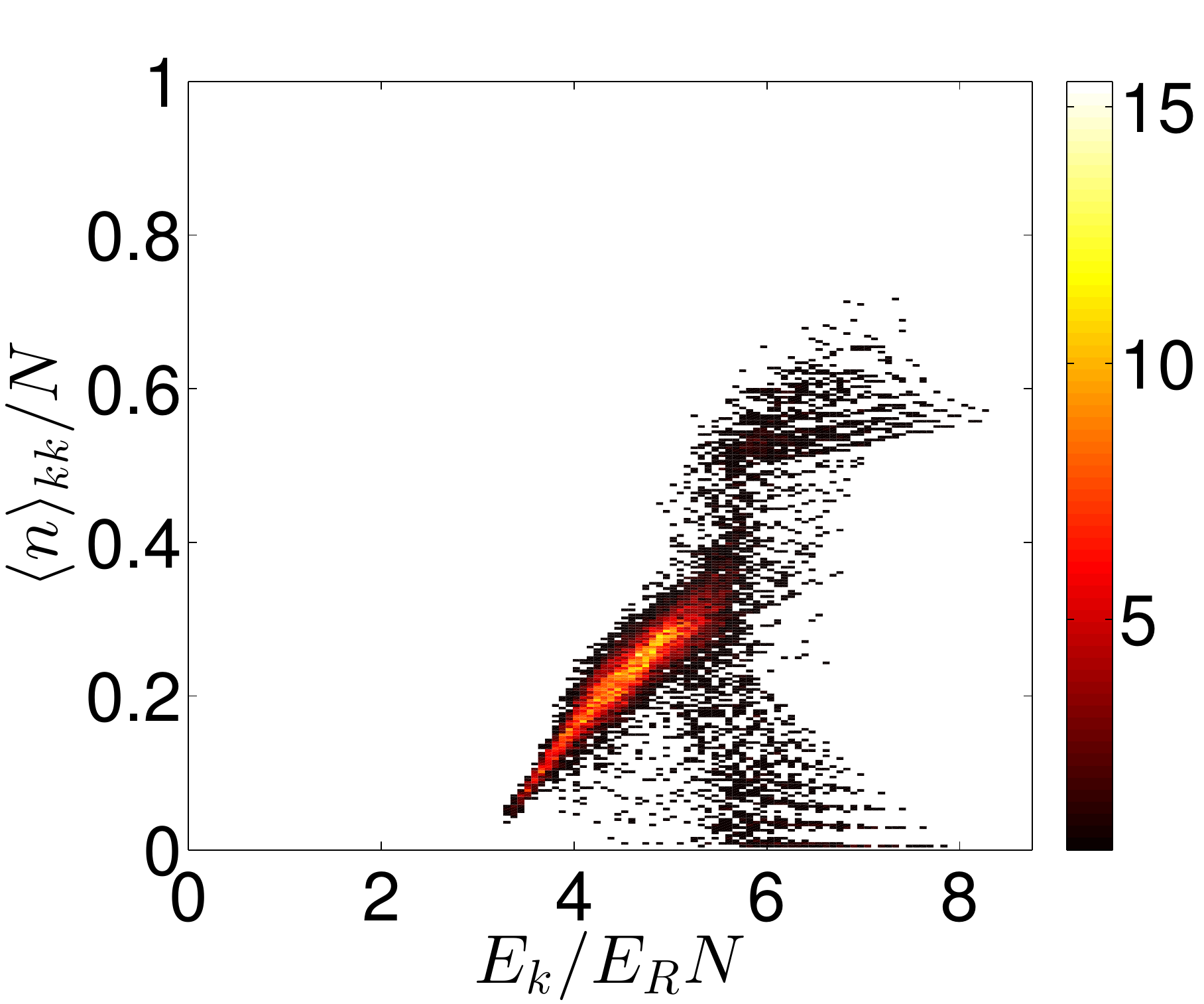}\includegraphics[width=0.5\columnwidth]{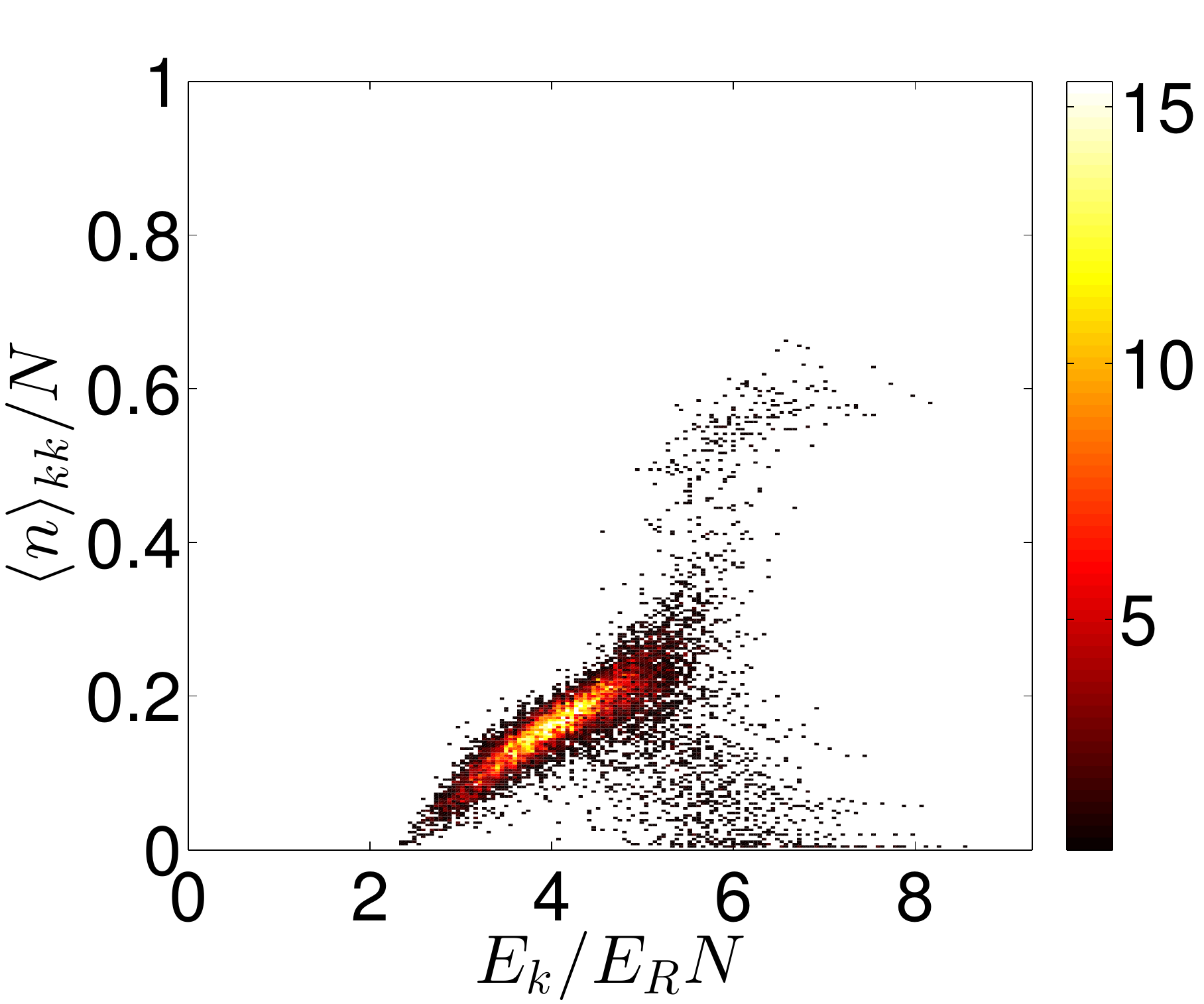}
\includegraphics[width=0.5\columnwidth]{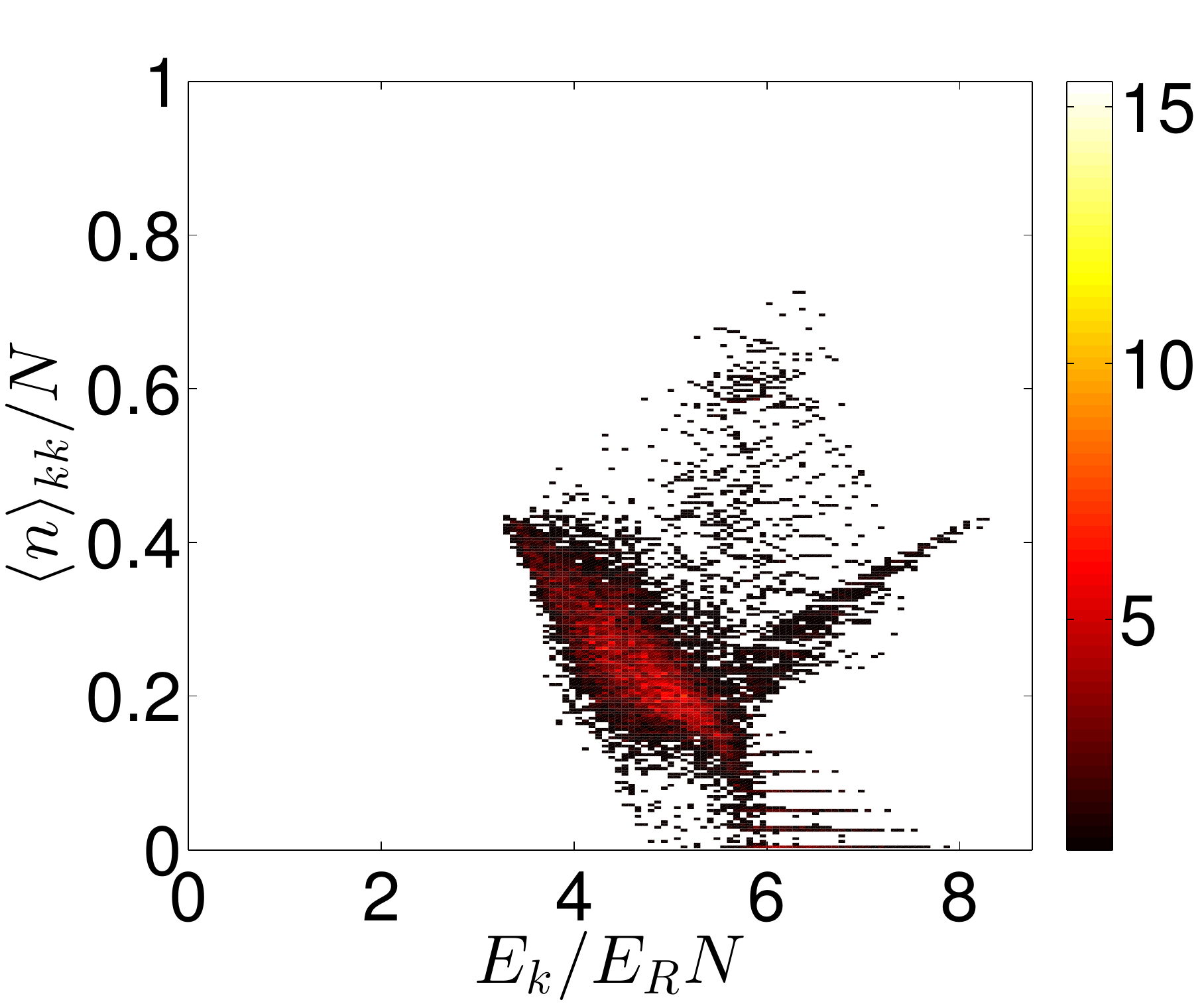}\includegraphics[width=0.5\columnwidth]{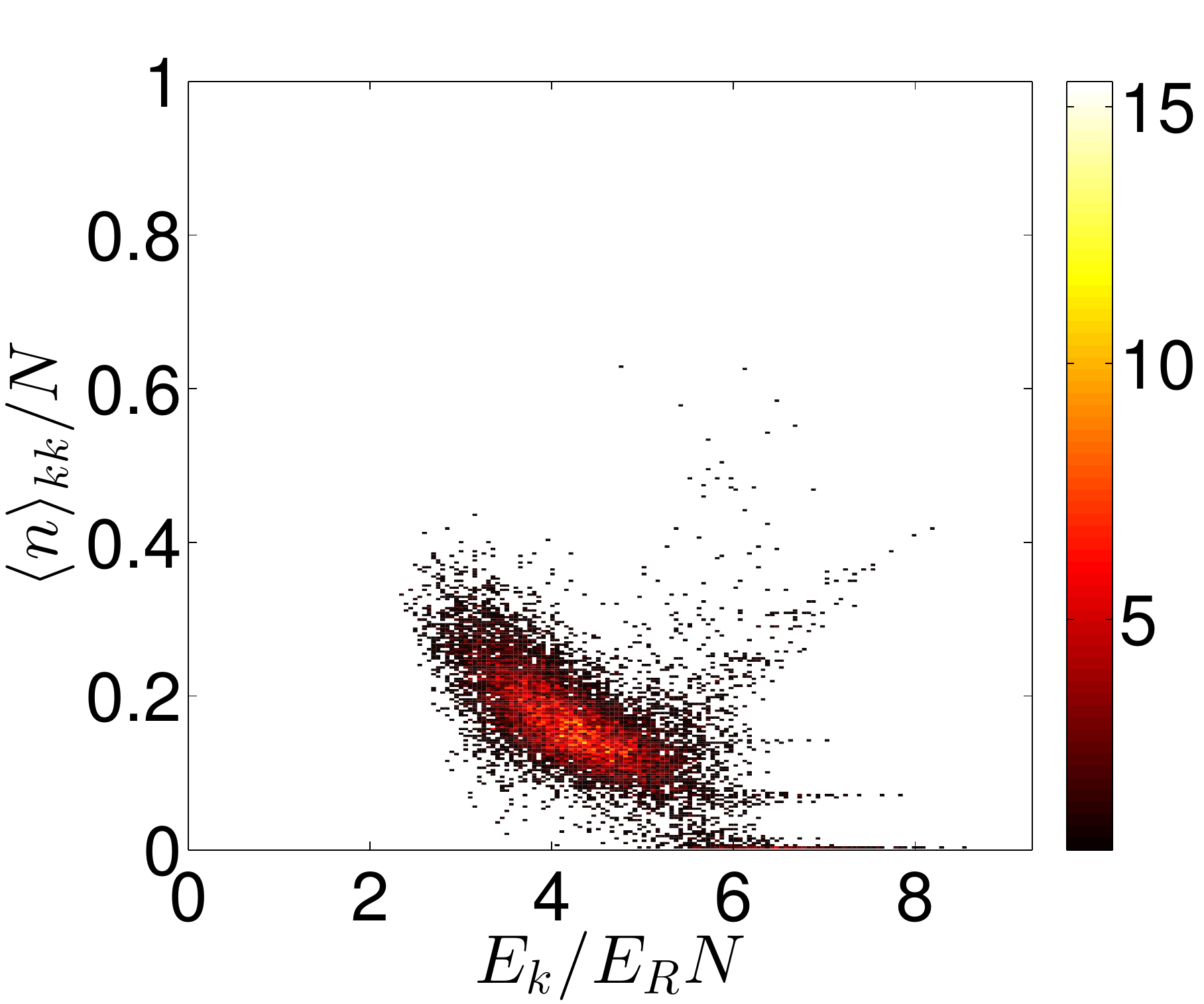}
\protect\caption{(Color online) Coarse-grained distribution of the EEV for $Ng/E_R=8$. (Top) upper-band manifold $M$ (Middle) upper-band of the second site $\{\ell=2,r=2\}$; (Bottom) lower-band of the second site $\{\ell=1,r=2\}$; (Left) $L=2$; and (Right) $L=3$.
}
\label{fig:eev}
\end{center}
\end{figure}

Following Refs. \cite{Cassidy2011,Kai2013}, we obtain the coarse-grained distribution of the EEV of the mode occupation number $\langle n \rangle _{kk}$. Examples of such distribution for observables when $Ng/E_R=8$ are depicted in Fig.~\ref{fig:eev} for lattice sizes $L=\{2,3\}$. As mentioned in the previous section, we expect the ETH to be satisfied for this choice of interaction strength both for $L=2$ and $L=3$. Indeed, ETH seems to work well for the upper-band occupation number $\langle \hat{n}^{2}_{r} \rangle$ and also for the manifold $\langle M \rangle=\langle \sum_r \hat{n}^{2}_r \rangle$, as evident from the bright regions in the first two rows of Fig.~\ref{fig:eev}. This is more obviously seen in energies close to the middle of the spectrum or the states close to the infinite-temperature limit where all the modes become equally populated. On the other hand, at least for the finite systems considered here, ETH appears to be violated for the lower-band occupation number 
$\langle \hat{n}^{1}_{r} \rangle$ as seen from the relative broadness of the EEV distribusion shown in the third row in Fig.~\ref{fig:eev}. Therefore, we do not expect to see the usual signatures of quantum thermalization in finite systems for the lower-band occupation numbers, i.e., the long-time limit of $\langle \hat{n}^1_r \rangle$ for different initial states are not expected to give the same value as the microcanonical ensemble prediction. Before we further elaborate on this observation, let us first introduce in the following discussion a more quantitative way of assessing the smoothness of the EEV distribution and in turn, the validity of ETH for a specific operator.

\begin{figure}[!ht]
\begin{center}
\includegraphics[width=0.5\columnwidth]{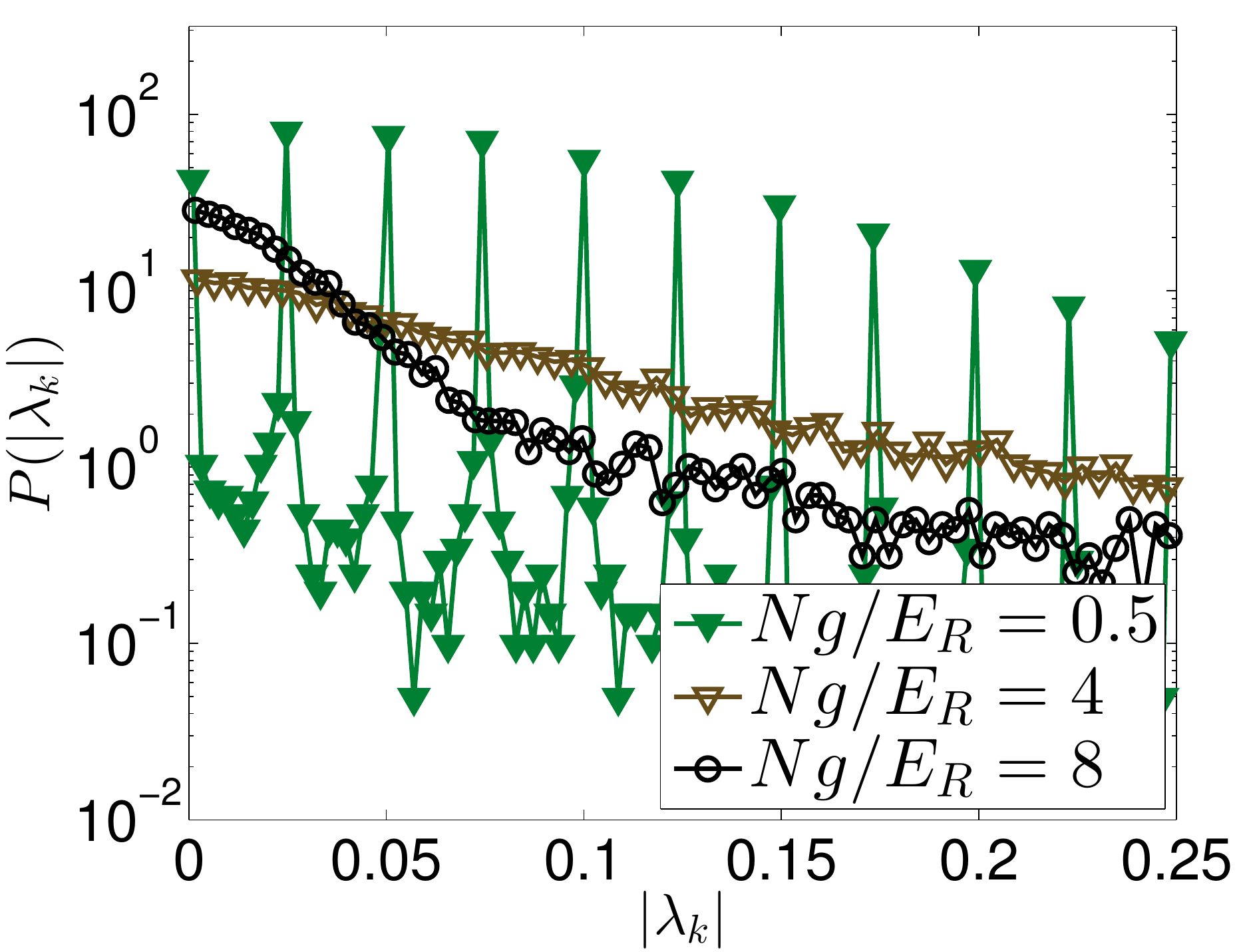}\includegraphics[width=0.5\columnwidth]{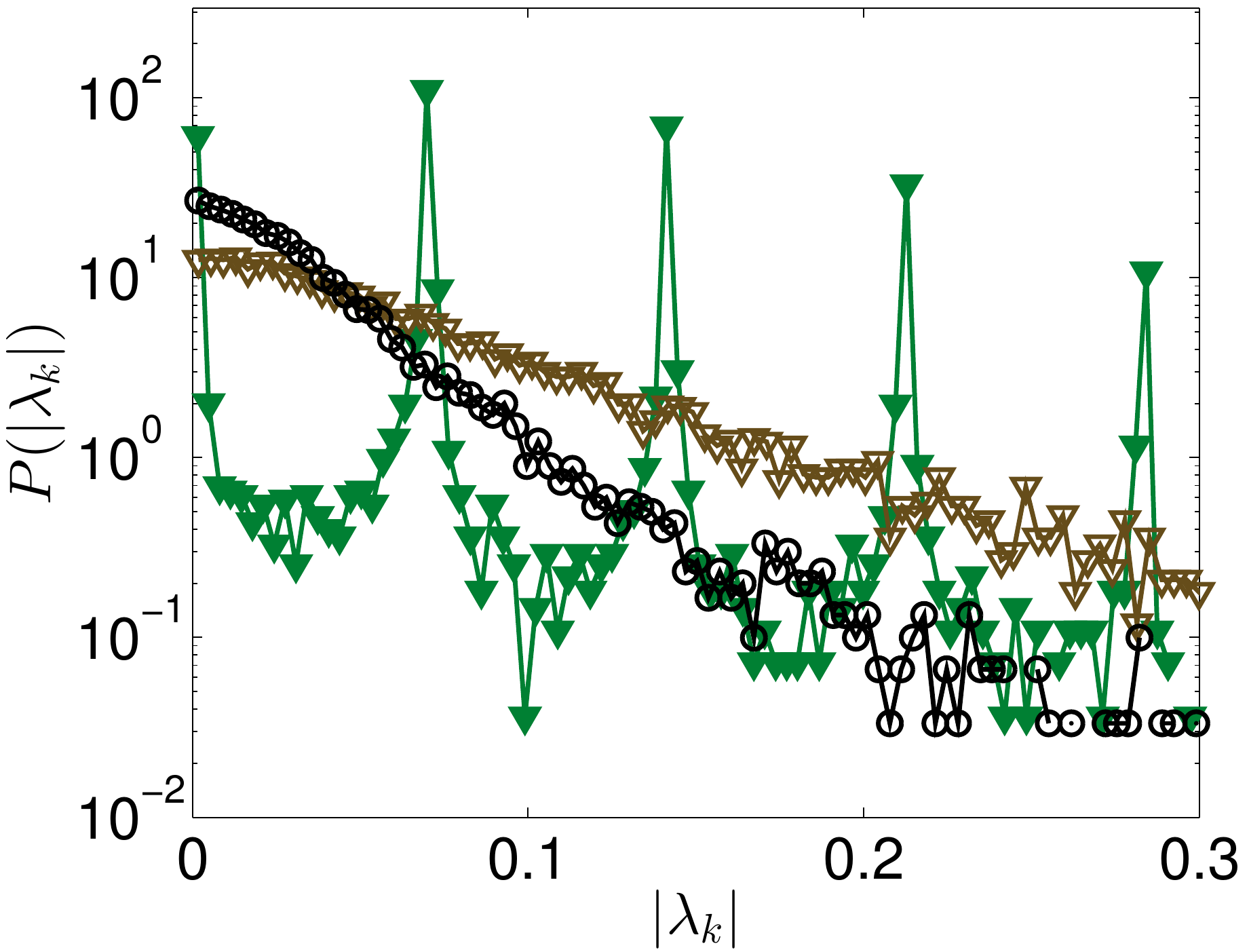}
\includegraphics[width=1.0\columnwidth]{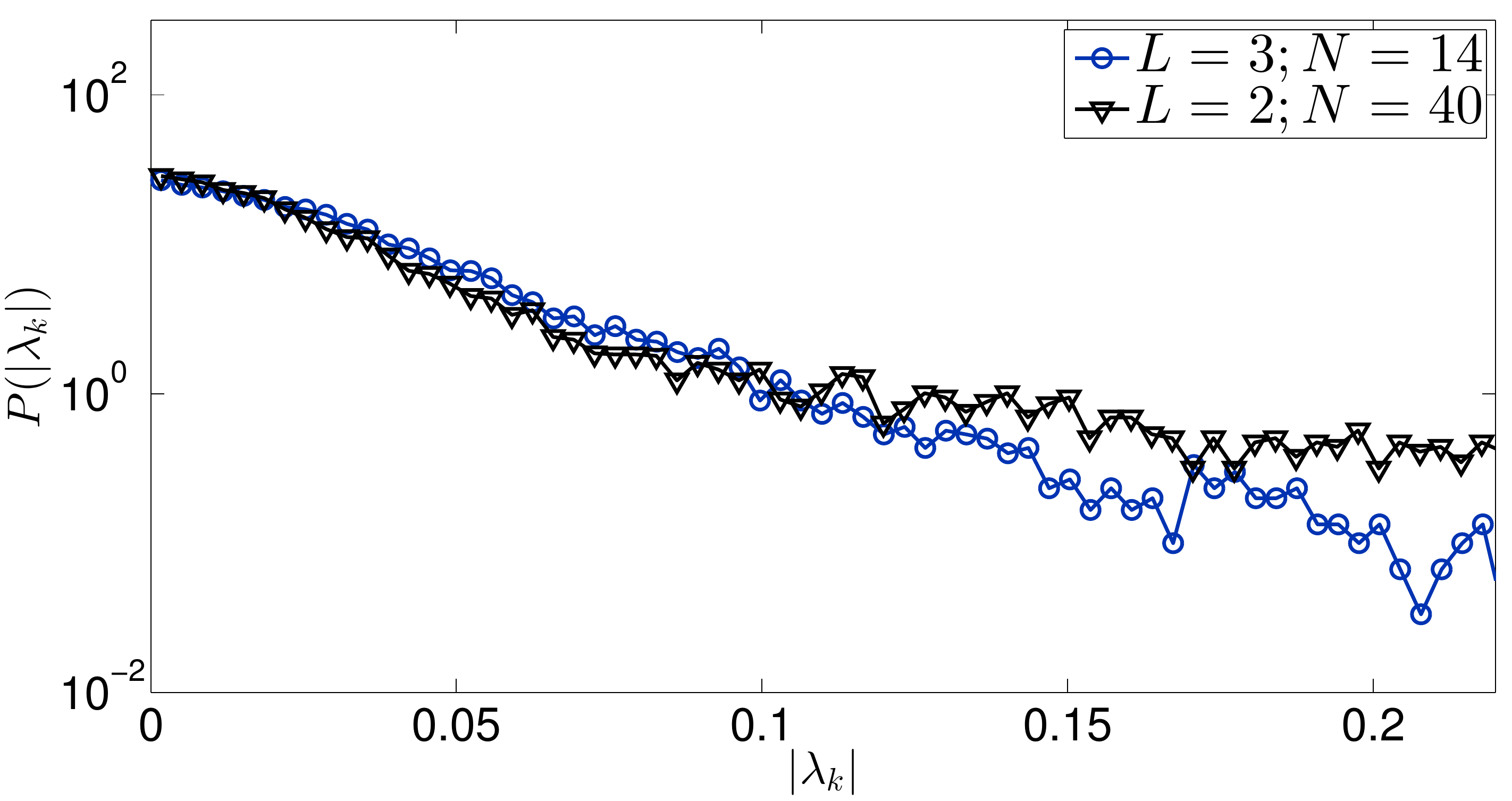}
\protect\caption{(Color online) Distributions of consecutive EEV gaps of the upper-band manifold $M$. (Top-Left) Comparison between different interaction couplings (Top-Left) for $L=2$ and (Top-Right) for $L=3$. (Bottom) comparison between $L=2$ and $L=3$ with fixed $Ng/E_R=8$.
}
\label{fig:eigdist}
\end{center}
\end{figure}

The fluctuation of the expectation values between the eigenstates of the system can be further analyzed using the distribution of gaps between consecutive EEV \cite{Kim2014,Cosme2015} given by,
\begin{equation}
 |\lambda_k| = \biggl| \langle M \rangle_{k+1,k+1} - \langle M \rangle_{k,k} \biggr|,
\end{equation}
for the manifold and 
\begin{equation}
 |\eta^{\ell}_{r,k}| = \biggl| \langle k+1| \hat{n}^{\ell}_{r} |k+1 \rangle - \langle k| \hat{n}^{\ell}_{r} |k \rangle \biggr|,
\end{equation}
for the mode-occupation numbers. The smoothness of the EEV as a function of the eigenenergies can be captured by the distribution $P(|\lambda_k|)$ and $P(|\eta^{\ell}_{r,k}|)$. That is, the distribution becomes sharply peaked around $|\lambda_k|=0$ or $|\eta^{\ell}_{r,k}|=0$ as the fluctuation becomes smoother. As a direct consequance, the support of $P(|\lambda_k|)$ or $P(|\eta^{\ell}_{r,k}|)$ decreases as the EEV smoothens. We shall only consider a bulk of the spectrum close to the middle by removing the lowest and the highest $10 \%$ of the eigenstates. 

\begin{figure}[!ht]
\begin{center}
\includegraphics[width=1.0\columnwidth]{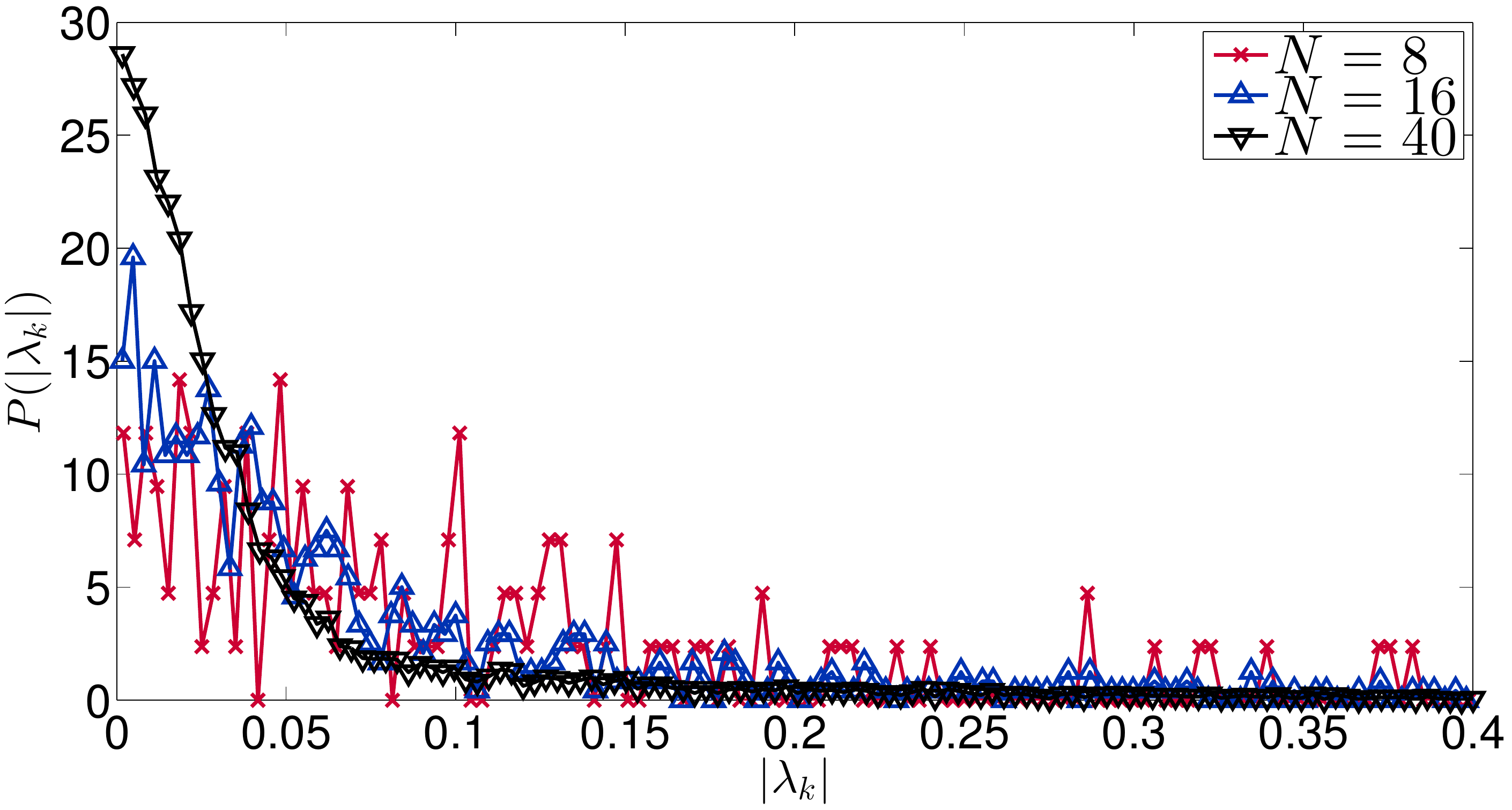}
\includegraphics[width=0.5\columnwidth]{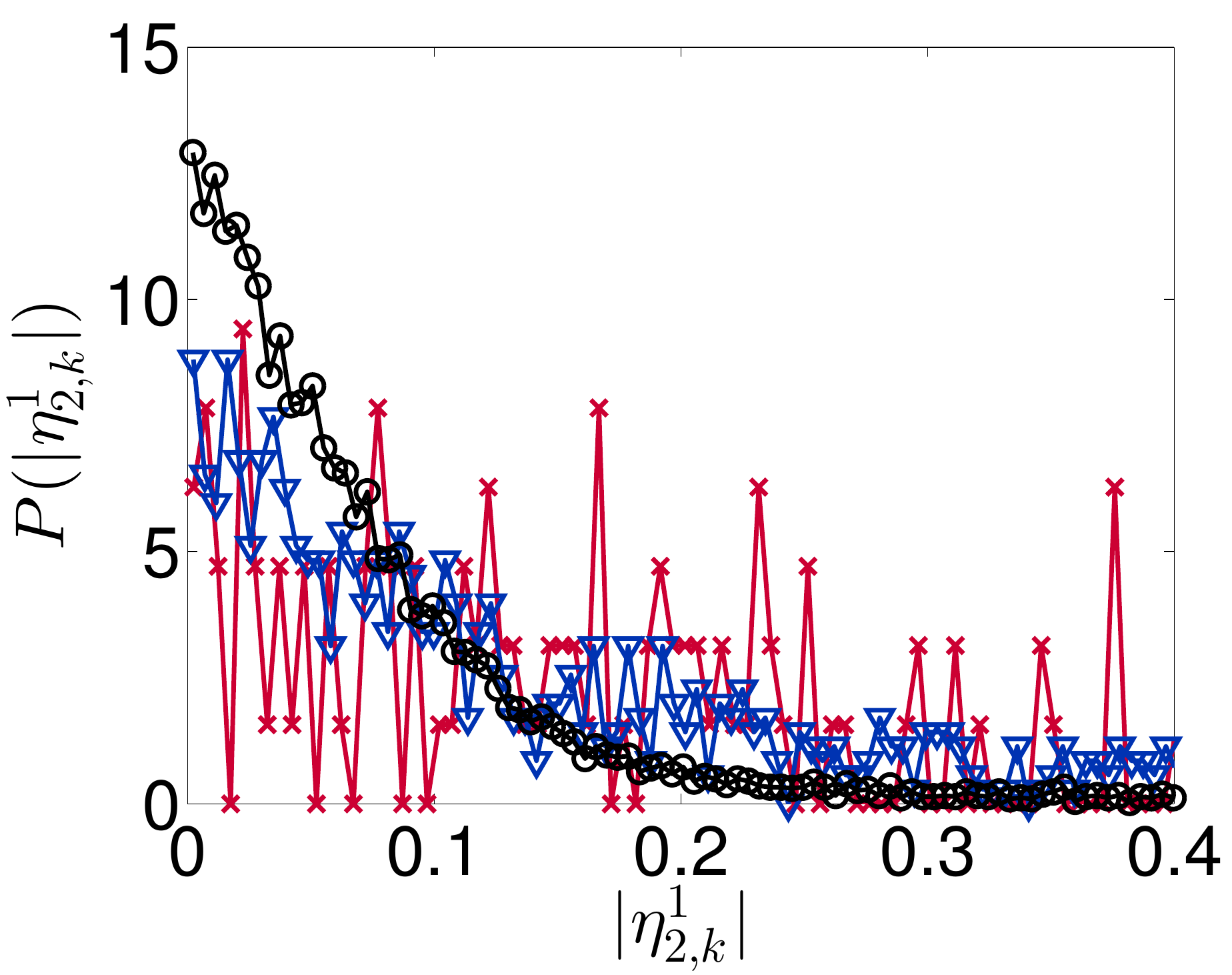}\includegraphics[width=0.5\columnwidth]{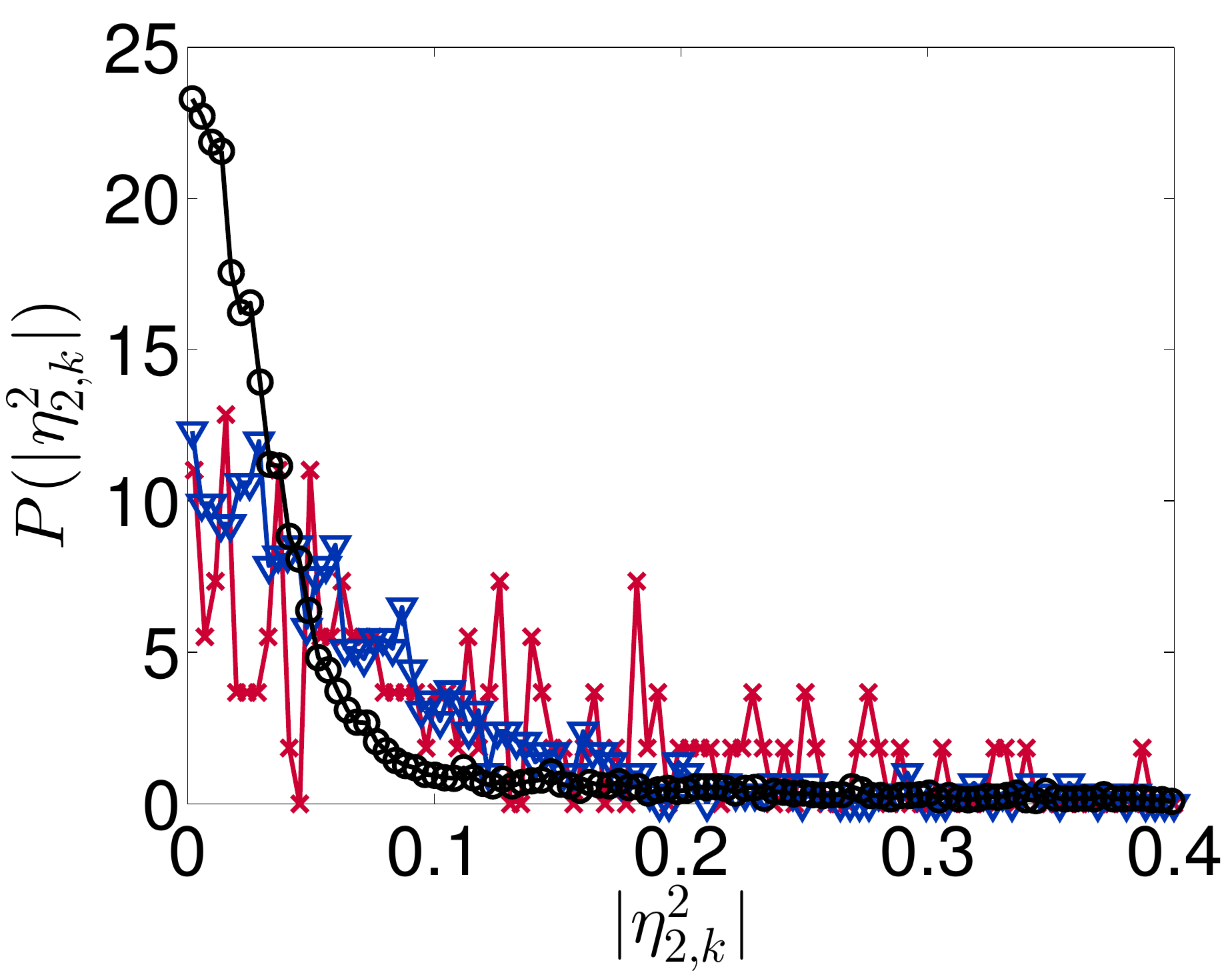}
\protect\caption{(Color online) Distributions of consecutive EEV gaps for $Ng/E_R=8$ and $L=2$ with varying number of bosons $N$. (Top) upper-band manifold $M$ (Middle) upper-band of the second site $\{\ell=2,r=2\}$; (Bottom) lower-band of the second site $\{\ell=1,r=2\}$;
}
\label{fig:eigdist2}
\end{center}
\end{figure}

In Fig.~\ref{fig:eigdist}, we plot $P(|\lambda_k|)$ for varying system sizes and interaction strengths. For weak interaction strength represented by $Ng/E_R=0.5$, peaks in the distribution $P(|\lambda_k|)$ appear for both system sizes $L=2$ and $L=3$. This can be explained by the fact that for sufficiently weak interaction strength the manifold or the total occupation in the upper-band (or the lower-band) becomes an almost good quantum number, which means horizontal bands emerge in the EEV distribution. Increasing the interaction strength up to $Ng/E_R=8$, smoothens the EEV and this makes the distribution $P(|\lambda_k|)$ more narrow. This behaviour can be seen in the upper panel of Fig.~\ref{fig:eigdist}. The support of the distribution also becomes narrower as the system size increase from $L=2$ and $L=3$ as shown in the lower panel of Fig.~\ref{fig:eigdist}. 

\begin{figure}[!ht]
\begin{center}
\includegraphics[width=0.46\columnwidth]{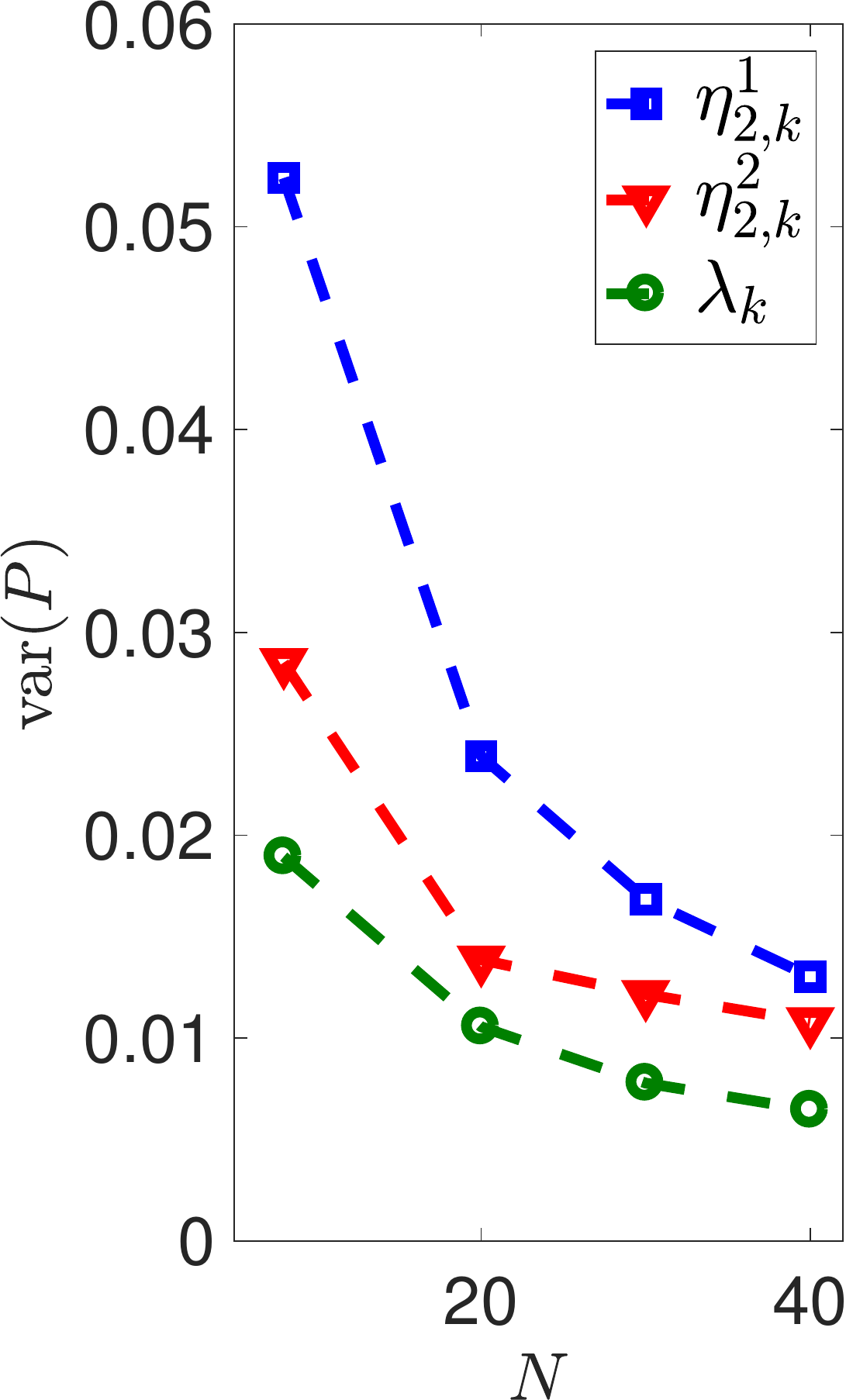}\hspace{0.1 cm}
\includegraphics[width=0.48\columnwidth]{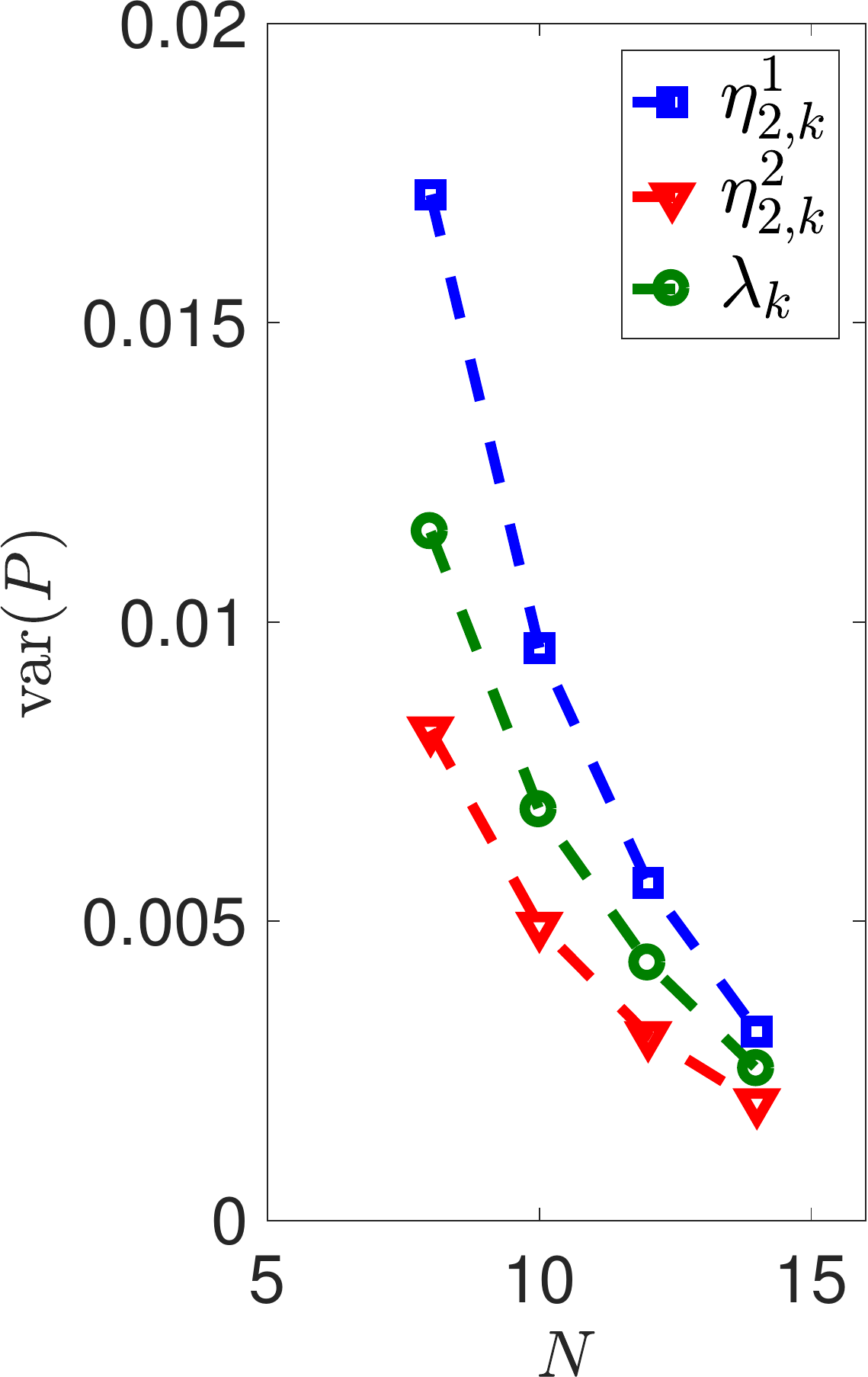}
\protect\caption{(Color online) Variances of the distributions of consecutive EEV gaps of different observables for varying $N$ and fixed $Ng/E_R=8$. (Left) $L=2$ and; (Right) $L=3$.
}
\label{fig:vareigdist}
\end{center}
\end{figure}

Let us now vary the number of particles $N$ and study how the distributions $P$ for both the manifold and the mode occupation numbers are affected. We show in Fig.~\ref{fig:eigdist2} the case when $Ng/E_R=8$ and $L=2$ for varying $N$. Here, we find that all the distributions become sharply peaked around zero as the semiclassical limit $N\to \infty$ is approached. In order to further quantify the rate at which the distributions become smoother, we have also calculated the corresponding variances of the distributions $P$ as we vary the number of bosons $N$ as shown in Fig.~\ref{fig:vareigdist}. 

Due to the exponential increase in the size of Hilbert space, we can only reach relatively low number of bosons for $L=3$ for this type of analysis. From the results presented in Fig.~\ref{fig:vareigdist}, the distribution for the lower-band mode occupation number $P(|\eta^{1}_{r,k}|)$ has consistently higher variance meaning it has broader distribution than the distributions for the upper-band modes and the manifold $P(|\eta^{2}_{r,k}|)$ and $P(|\lambda_k|)$, respectively. This observation reflects the behavior of apparent ETH violation for observables in the lower band as we have already seen from the coarse-grained distribution in the third column of Fig.~\ref{fig:eev}. Nevertheless, it is encouraging to notice that there is an apparent decrease in the variances of the distributions as $N$ becomes large not only for observables in the upper-band but also for those in the lower bands. This means that while quantum thermalization may be absent within the lower-band observables for finite system sizes, as already mentioned before, there are still signs that some form of relaxation may eventually arise in the semiclassical limit of large $N$. This observation justifies the investigation of possible relaxation dynamics within the mode occupation numbers in the large $N$-limit as discussed in the next section.

\section{Relaxation dynamics in the semiclassical limit}\label{sec:equil}

In the following numerical experiments, we consider initial product states in each mode with average energy around the region where the EEV for the manifold is smooth and concentrated. We specifically choose initial states with mean energy around $\langle \hat{H} \rangle/E_{R} N \approx 4$. More importantly, for the remainder of this work, we consider large filling factors $N/\mathcal{M} \gg 1$ and semiclassical limit of large $N \sim 10^4$. These conditions make the quantum dynamics amenable to the TWA.

We proceed to demonstrate the equilibration process starting from two different kinds of initial state: (a) coherent state and (b) Fock state. We focus on the two-site ($L=2$) and the three-site ($L=3$) models. The two-site system portrays one limiting case with only one tunneling term in Eq.~\eqref{eq:eom}. The three-site system is a minimal model that allows for atoms in one site (the central site $r=2$) to tunnel in both directions, i.e., first ($r=1$) and third ($r=3$) sites. We use $10^4$ trajectories in the TWA simulations for the two-site system with $N=4.0 \times 10^4$. For consistency with the exact diagonalization results for $N=14$ and $L=3$ shown in Fig.~\ref{fig:eev}, we use $N=1.4 \times 10^4$ and $5 \times 10^3$ trajectories for the three-site model.

\begin{figure}[!ht]
\begin{center}
\includegraphics[width=1.0\columnwidth]{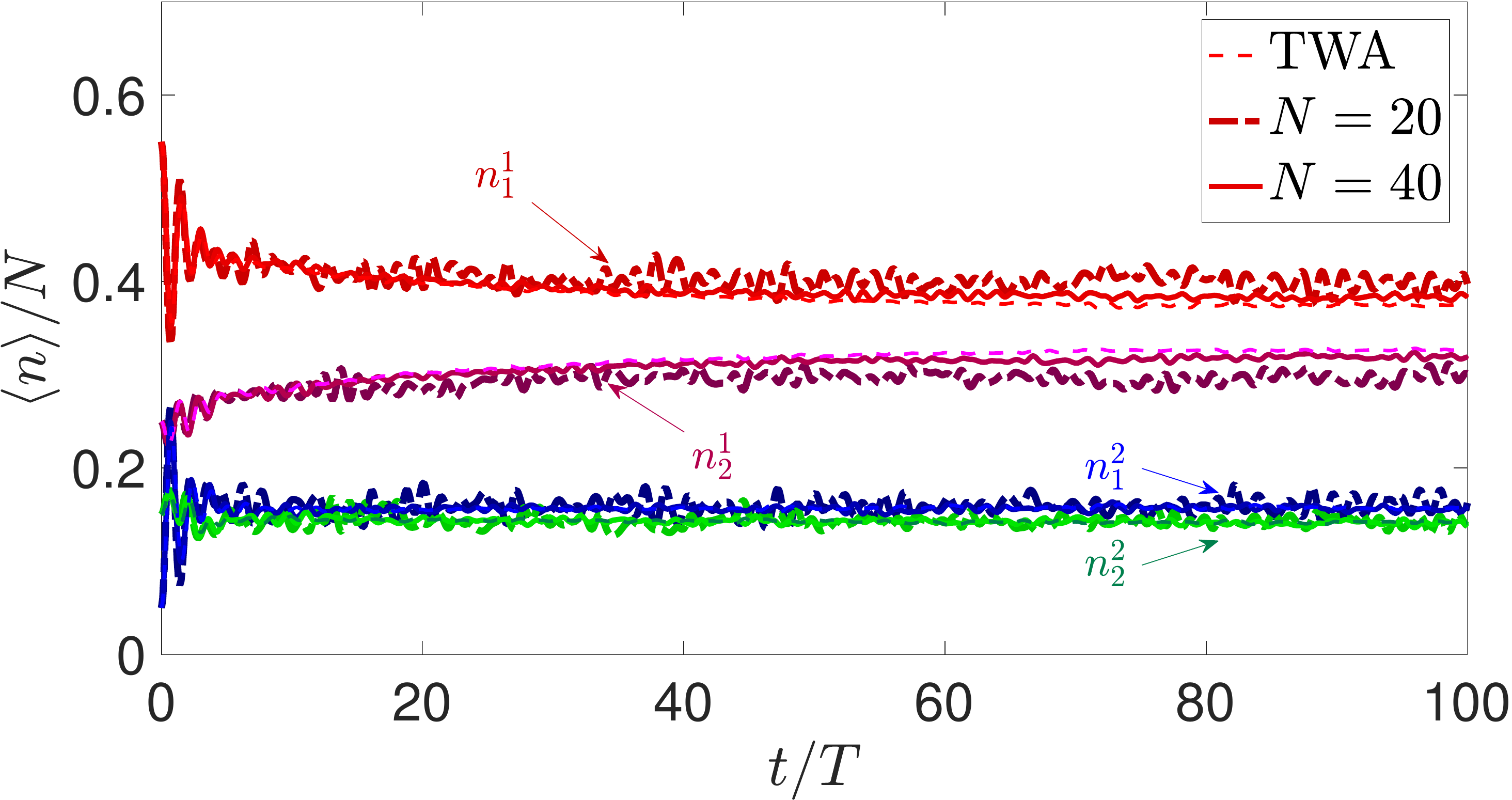}
\includegraphics[width=1.0\columnwidth]{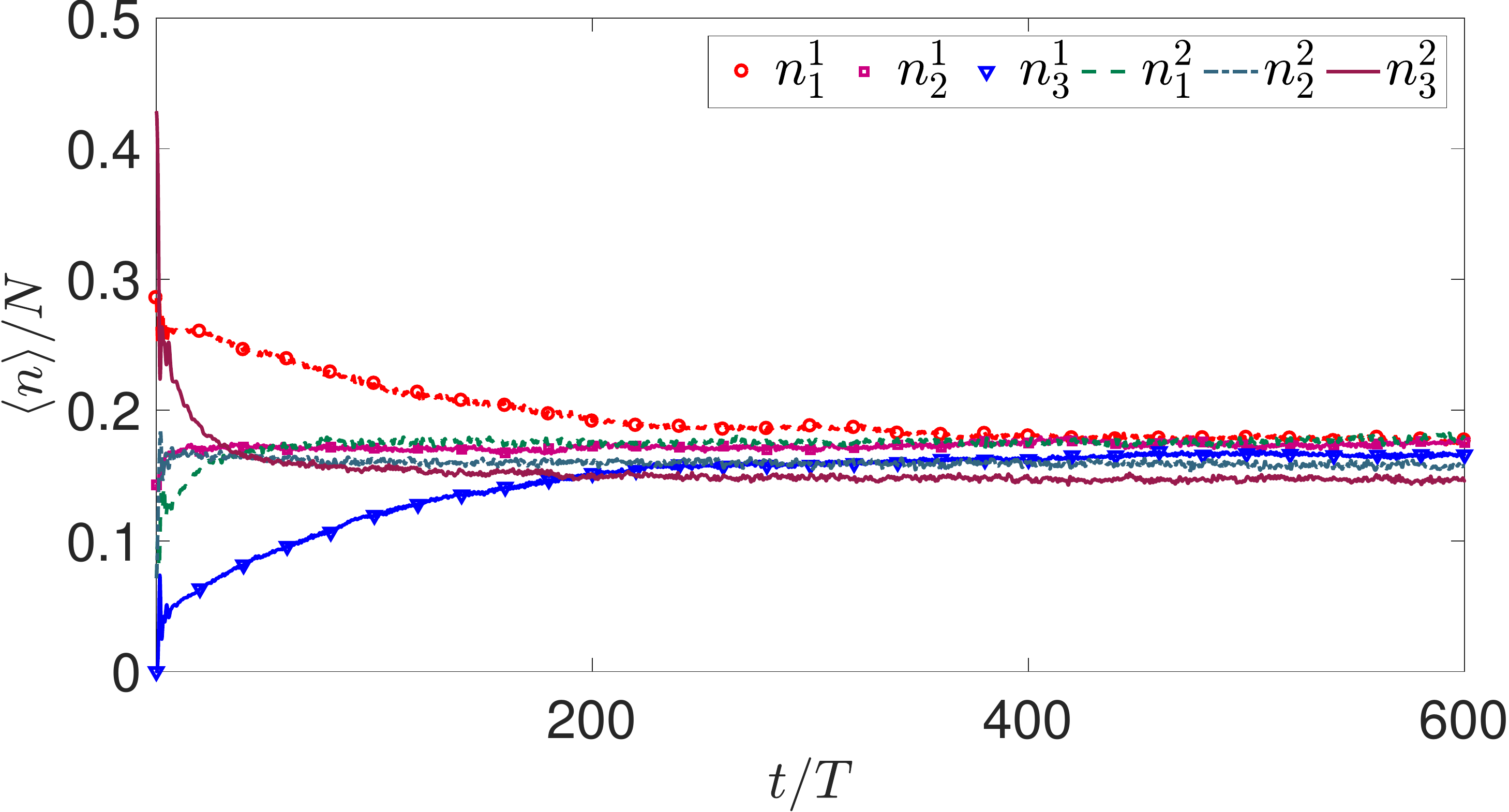}
\protect\caption{(Color online) Examples of dynamics of the mode occupation number with initial Fock states. (Top) Comparison between exact quantum dynamics (ED) and TWA simulation for $L=2$, $Ng/E_R=8$, and initial state of $\{n^1_1,n^1_2,n^2_1,n^2_2\}/N = \{0.55,0.25,0.05,0.15\} $; (Bottom) TWA results for the dynamics of mode occupation number for $L=3$, $Ng/E_R=8$, and initial state of $\{n^1_1,n^1_2,n^1_3,n^2_1,n^2_2,n^2_3\} = \{4,2,0,1,1,6\}\times 10^4 $.}
\label{fig:modes}
\end{center}
\end{figure}

To demonstrate equilibration of local observables, we investigate the time evolution of the mode occupation number $\langle \hat{n}^{\ell}_r \rangle$. We present examples of the relaxation dynamics of the mode occupation numbers in Fig.~\ref{fig:modes}. For large filling factors in the case of $L=2$, we find good agreement between the exact quantum dynamics ($N=40$) and the results of TWA ($N=10^4$) as shown in Fig.~\ref{fig:modes}. It also worth mentioning that the apparent absence of revivals within the timescale explored here is a consequence of the large $N$ limit. Unfortunately for $L=3$, we can not get a fair comparison between exact quantum dynamical and TWA results because the largest filling factor accessible in our quantum mechanical simulations is only $N/\mathcal{M}=2.333$. This is a relatively low number when compared to filling factor for $L=2$ ($N/\mathcal{M}=10$) where we find good agreement between the two methods as seen in upper panel of Fig.~\ref{fig:modes}. Nevertheless, we expect the system to exhibit thermalization dynamics in the limit of large $N$ as evident from the results shown in the right panel of Fig.~\ref{fig:vareigdist}.
We also present in Fig.~\ref{fig:modes} an example of relaxation when $L=3$, which exhibits plateaus in the dynamics of $\langle \hat{n}^{\ell}_r \rangle$ for sufficiently long time scales.

One hallmark of thermalization is the independence of a stationary state from details of initial states apart from the conserved quantities in the system. 
\begin{figure}[!ht]
\begin{center}
\includegraphics[width=1.0\columnwidth]{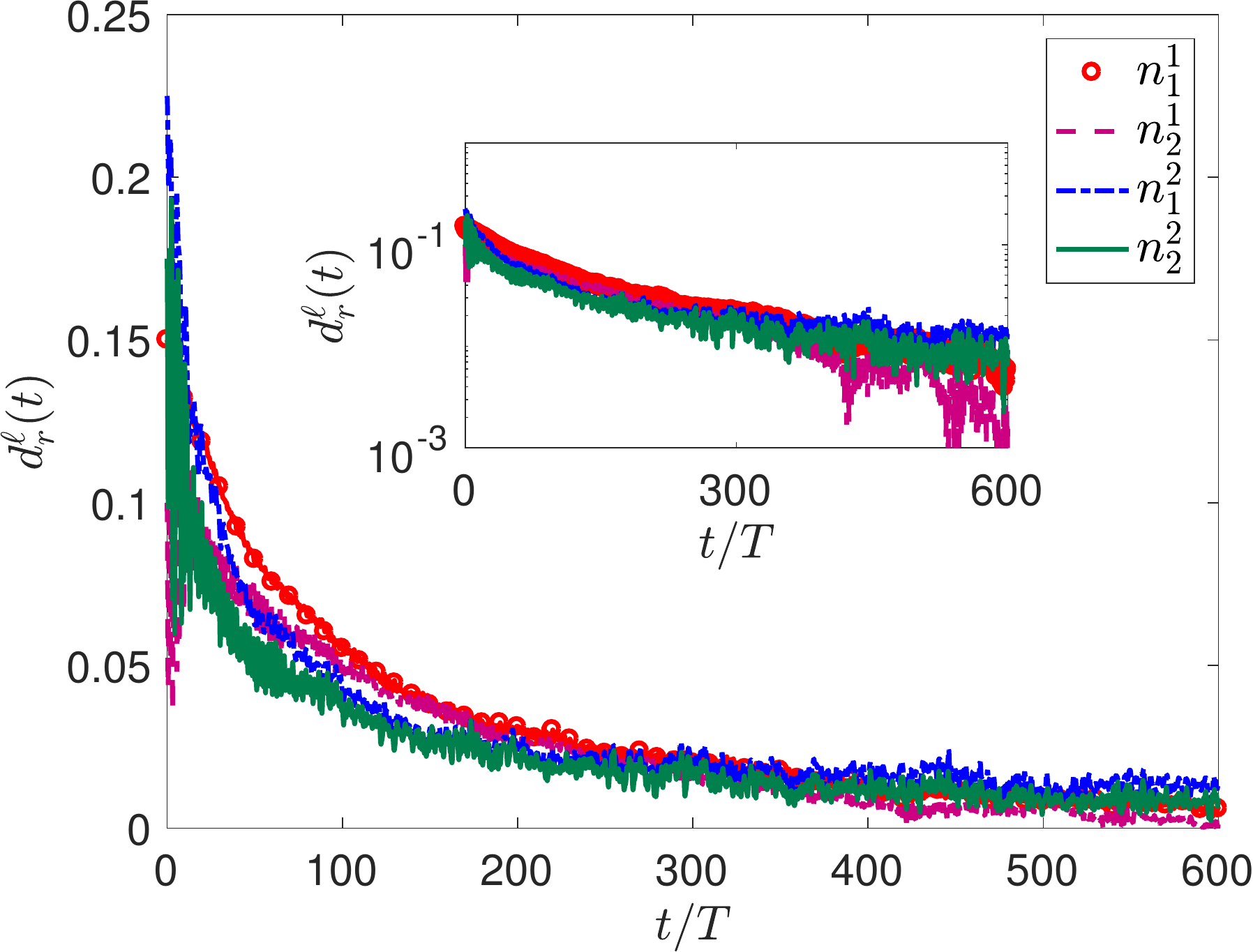}\\\includegraphics[width=1.0\columnwidth]{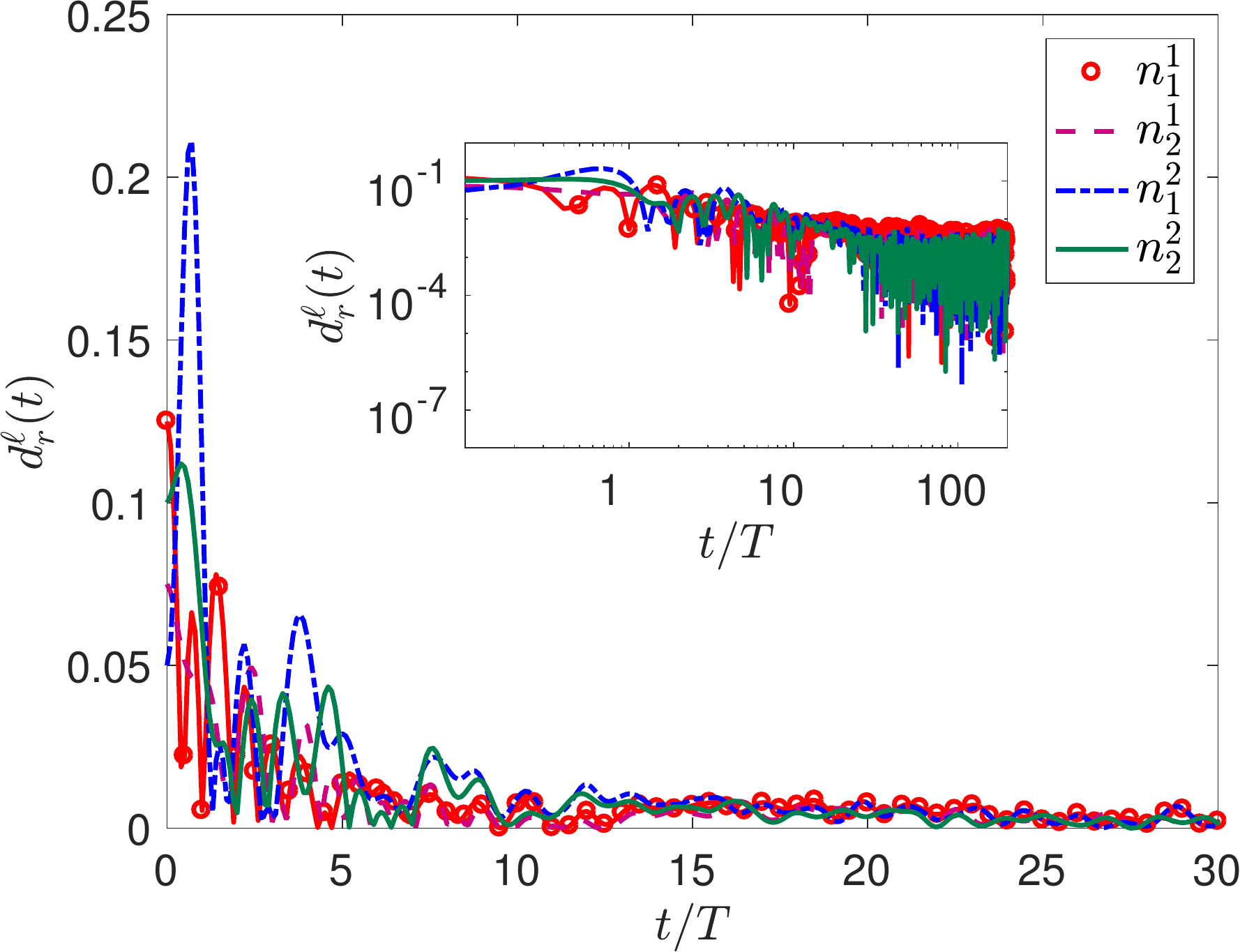}
\protect\caption{(Color online) Time evolution of the distance $d^{\ell}_r(t)$ between two different initial Fock states with the same energies for $L=2$ (Top) $Ng/E_R=4$ with initial states $\{\psi_1\}=\{n^1_1,n^1_2,n^2_1,n^2_2\} = \{0,14,13,13\}\times 10^4 $ and $\{\psi_2\}=\{6,10,4,20\}\times 10^4 $ and (Bottom) $Ng/E_R=8$ with initial states $\{\psi_1\}= \{23,10,2,5\}\times 10^4 $ and $\{\psi_2\}=\{18,13,0,9\}\times 10^4 $. Inset: (Top) Semilogarithmic scale in $d^{\ell}_r(t)$. (Top) Log-log scale in $d^{\ell}_r(t)$ and $t$.}
\label{fig:focks2}
\end{center}
\end{figure}
\begin{figure}[!ht]
\begin{center}
\includegraphics[width=1.0\columnwidth]{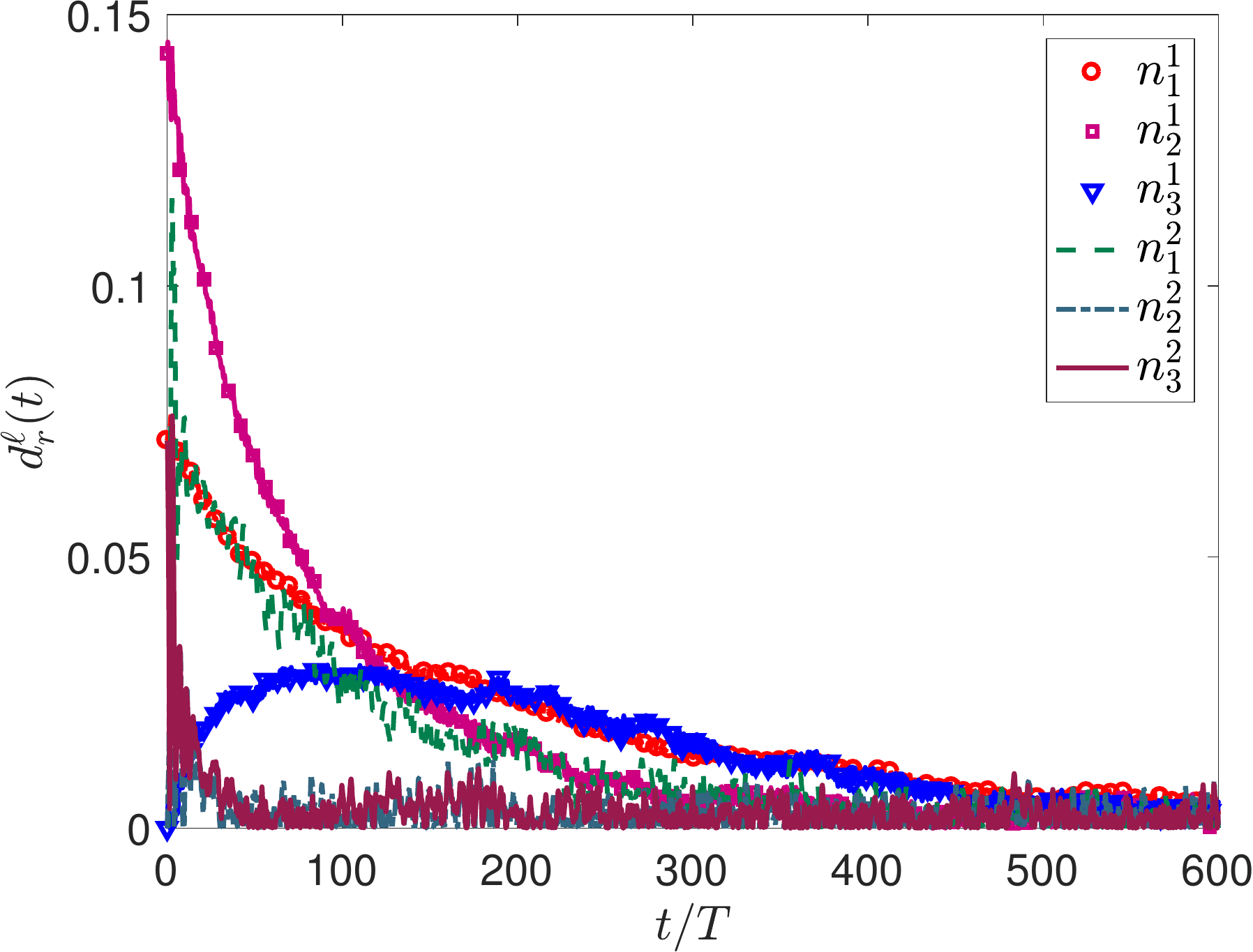}\\\includegraphics[width=1.0\columnwidth]{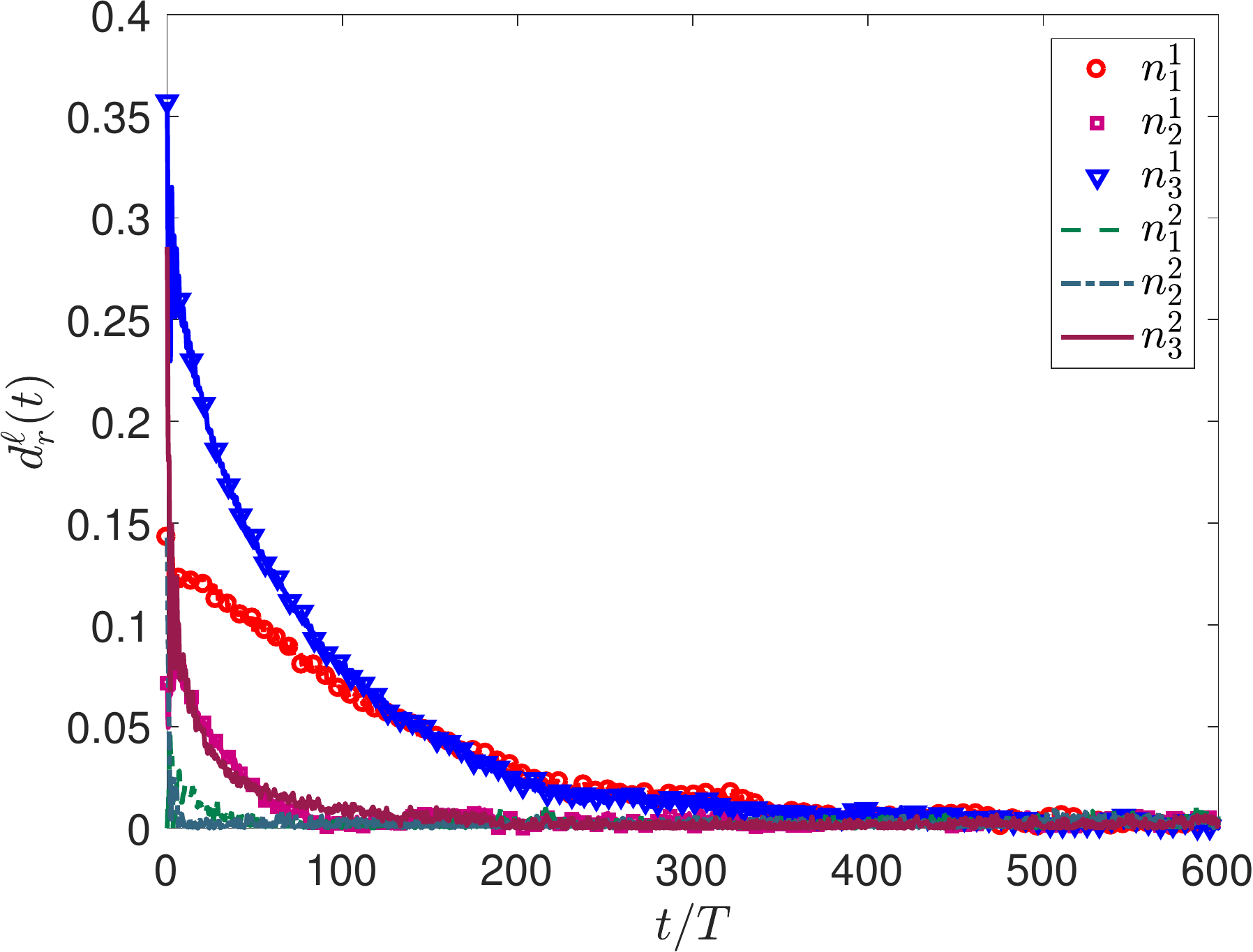}
\protect\caption{(Color online) Time evolution of the distance $d^{\ell}_r(t)$ between two different initial Fock states with the same energies for $L=3$ (Top) $Ng/E_R=4$ with initial states $\{\psi_1\}=\{n^1_1,n^1_2,n^1_3,n^2_1,n^2_2,n^2_3\} = \{1,1,1,2,5,4\}\times 10^4 $ and $\{\psi_2\}=\{0,3,1,2,5,3\}\times 10^4 $ and (Bottom) $Ng/E_R=8$ with initial states $\{\psi_1\}= \{4,2,0,1,1,6\}\times 10^4 $ and $\{\psi_2\}=\{2,1,5,1,3,2\}\times 10^4 $.}
\label{fig:focks3}
\end{center}
\end{figure}
Suppose that we choose two different initial states with very similar initial energy, $\langle \hat{H} \rangle_{1} \approx \langle \hat{H} \rangle_{2} \approx E_0$, and the same total number of bosons. Then, a system can be considered to have properly thermalized if in the long-time limit, the distance between two initial states measured by
\begin{equation}
	d^{\ell}_r(t) = \biggl|\langle \hat{n}^{\ell}_r(t) \rangle_1 - \langle \hat{n}^{\ell}_r(t) \rangle_2\biggr|
\end{equation}
approaches zero. Physically, this means that the two initial states dynamically approach the same steady-state for the mode occupation numbers. This quantity is equivalent to the dynamical distance used in Ref.~\cite{Nessi2015} to examine prethermalization dynamics in a system of fermions after a quantum quench. 

\begin{figure}[!ht]
\begin{center}
\includegraphics[width=1.0\columnwidth]{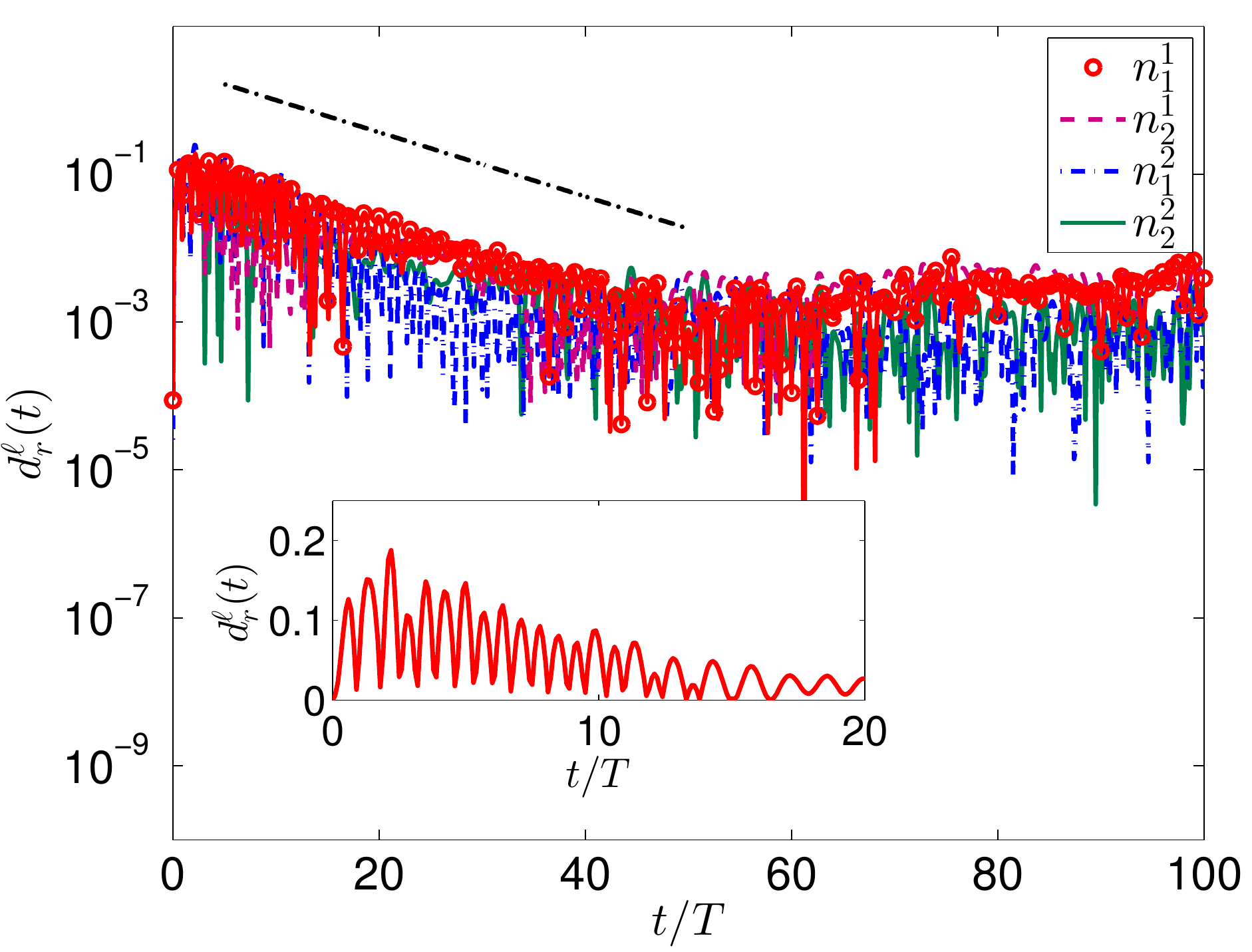}
\includegraphics[width=1.0\columnwidth]{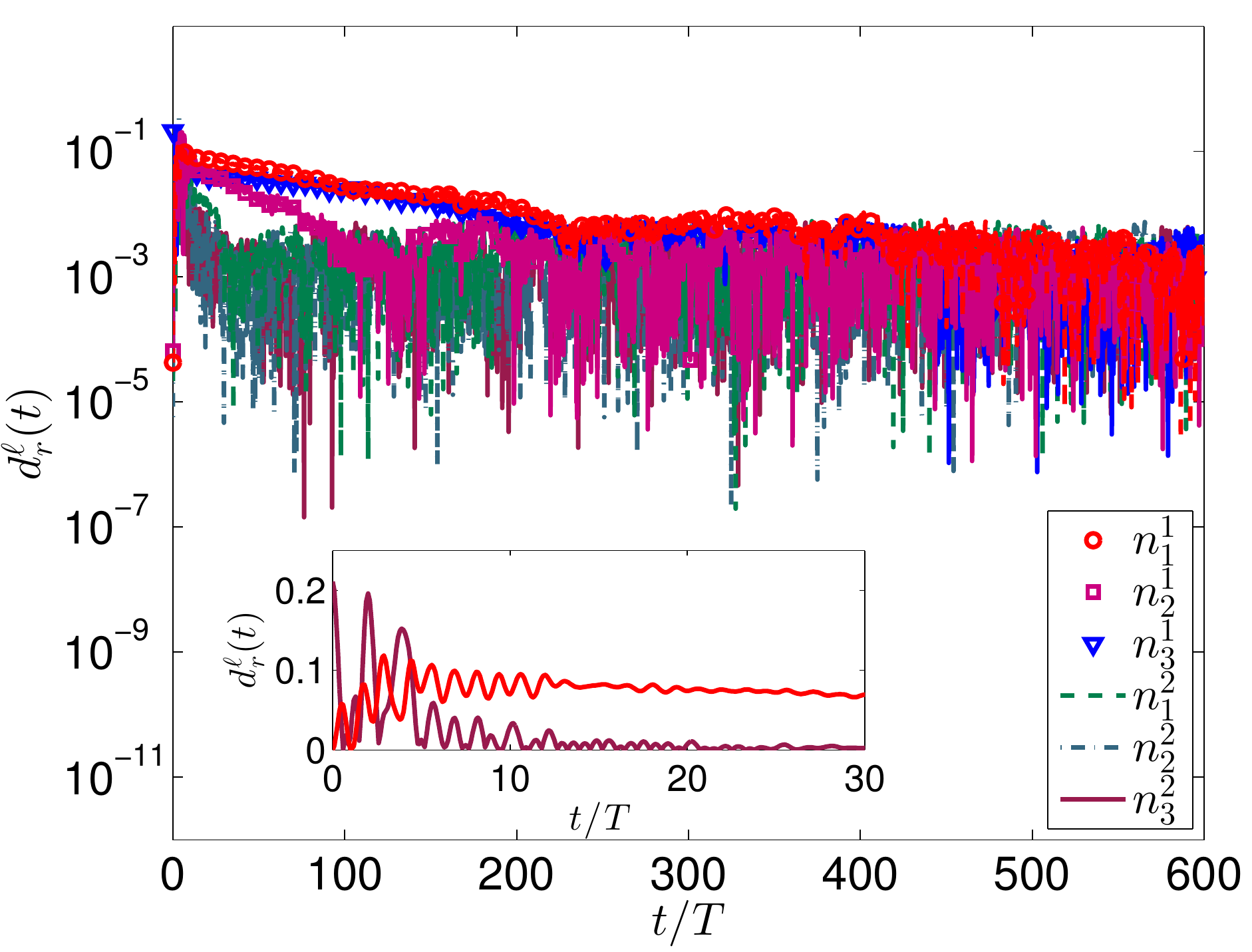}
\protect\caption{(Color online) Time evolution of the distance $d^{\ell}_r(t)$ between an initial Fock and coherent states with the same initial energies for $Ng/E_R=8$; (Top) $L=2$ with Fock initial state of $\{\psi_1\}=\{n^1_1,n^1_2,n^2_1,n^2_2\} = \{23,10,2,5\}\times 10^4 $ and coherent initial state of $\{\psi_2\}=\{23,13,2,2\}\times 10^4 $; and (Bottom) $L=3$ with Fock initial state of $\{\psi_1\}=\{n^1_1,n^1_2,n^1_3,n^2_1,n^2_2,n^2_3\} = \{4,2,0,1,1,6\}\times 10^4 $ and coherent initial state of $\{\psi_2\}=\{4,2,3,1,1,3\}\times 10^4 $. The black dashed-dotted line represents an exponential decay as a guide to the eye. Inset: (Top) Short-time oscillations of $d^{\ell}_r(t)$. (Bottom) Comparison between the short-time decay of $d^{\ell}_r(t)$ in $\langle \hat{n}^1_1 \rangle$ and $\langle \hat{n}^2_3 \rangle$.}
\label{fig:fockcohe2}
\end{center}
\end{figure}
\begin{figure}[!ht]
\begin{center}
\includegraphics[width=0.9\columnwidth]{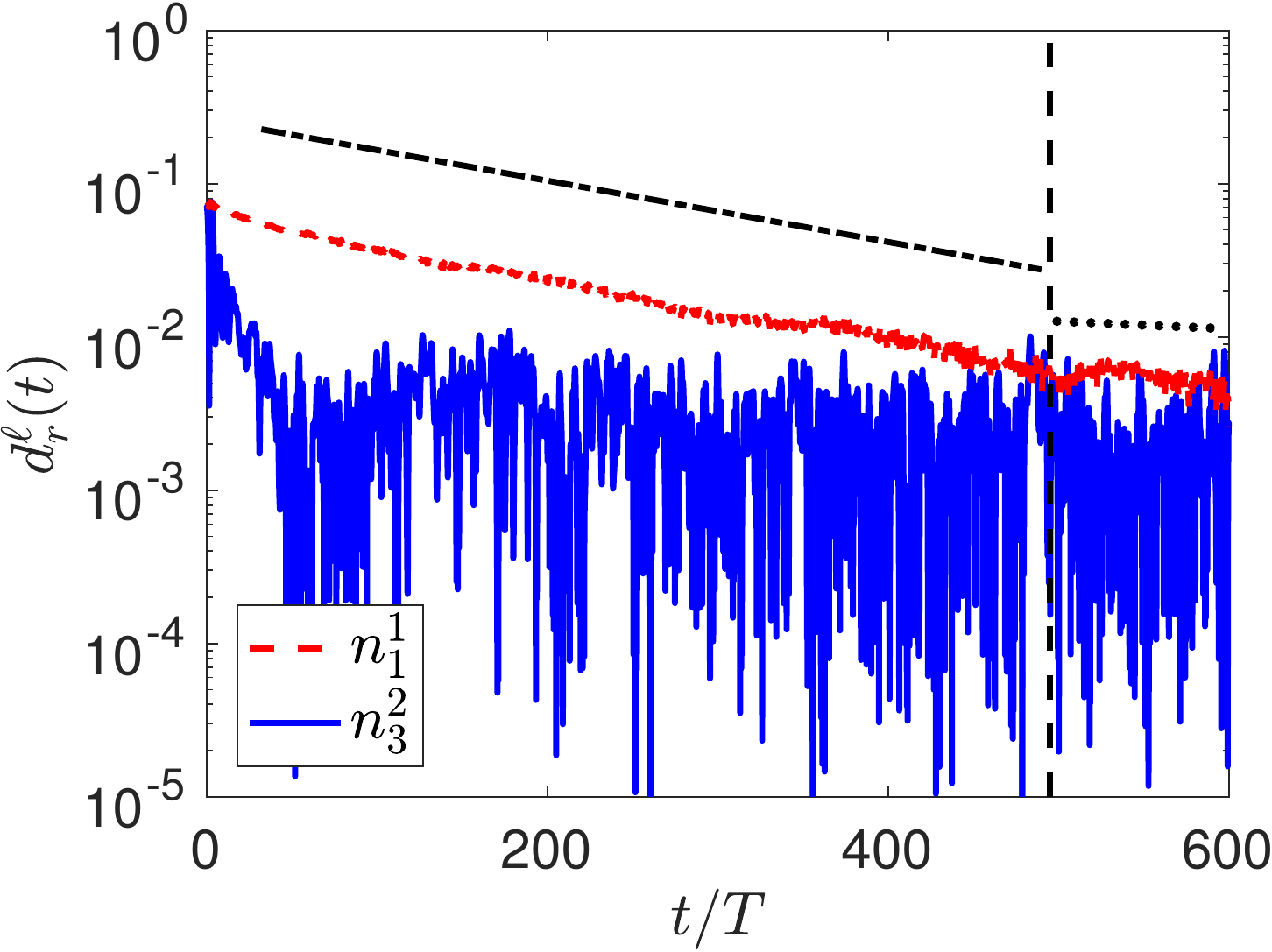}
\includegraphics[width=0.9\columnwidth]{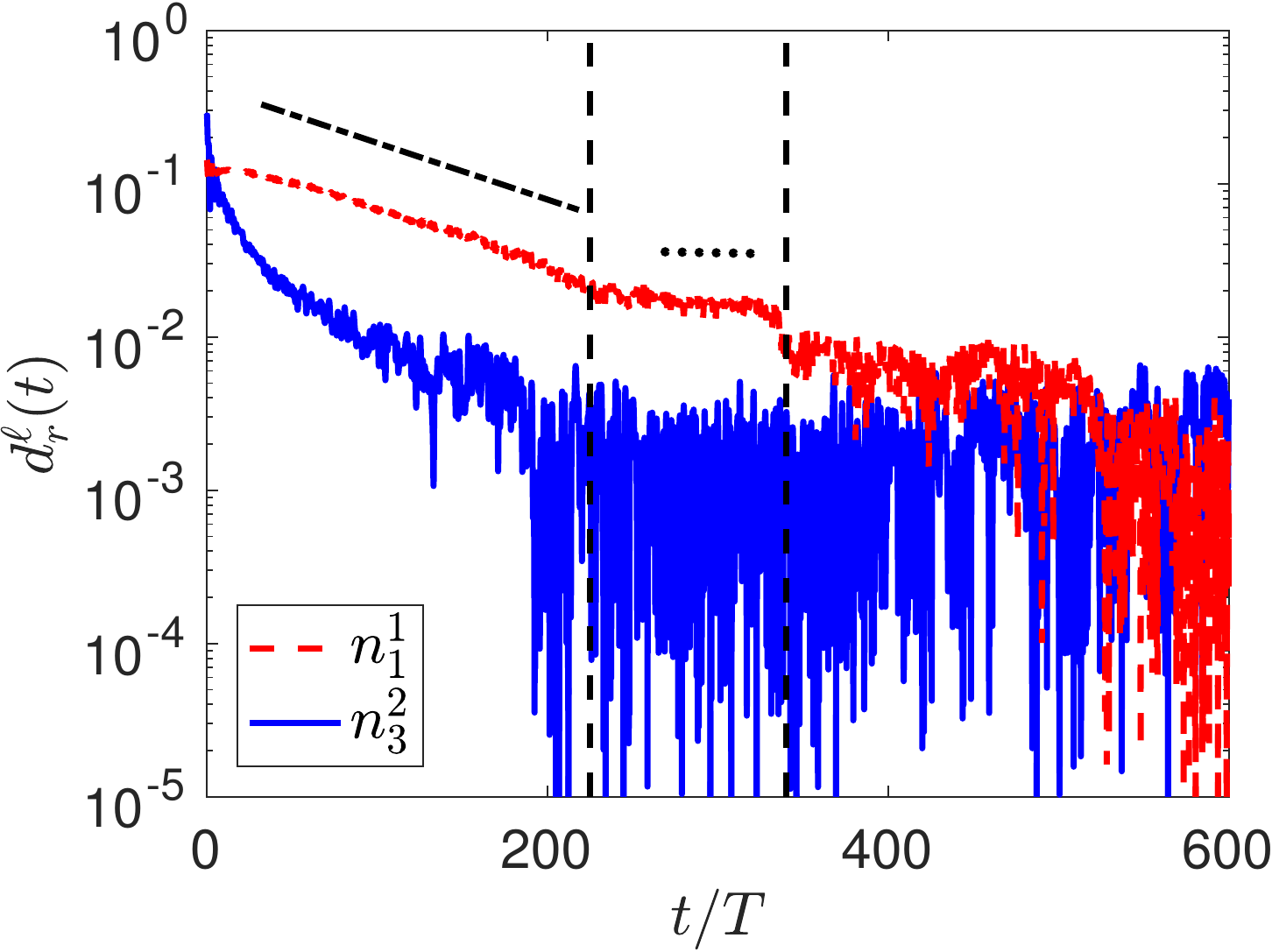}
\protect\caption{(Color online) Time evolution of the distance, $d^{\ell}_r(t)$, in semi-logarithmic scale between two different initial Fock states with the same initial energies for $L=3$ as in Fig.~\ref{fig:focks3}. (Top) $Ng/E_R=4$ and (Bottom) $Ng/E_R=8$. The black broken lines correspond to exponential fits of the form $\sim \mathrm{exp} (-t/\tau)$. The dotted lines emphasize the prethermalization plateaus.}
\label{fig:logdis3}
\end{center}
\end{figure}

In Fig.~\ref{fig:focks2}, we present the time evolution of the dynamical distance between two initial Fock states for $L=2$.  We observe fast thermalization in all the modes of the fully quantum chaotic case, $Ng/E_R=8$. However, for $Ng/E_R=4$, the system appears to be nonergodic as evident from the fact that the $d^{\ell}_r(t)$ never completely reaches zero in this case. This is clearly seen from comparing the inset plots in Fig.~\ref{fig:focks2}, where $d^{\ell}_r(t)$ fluctuates around a set of values that are several orders of magnitude smaller in $Ng/E_R=8$ than that in $Ng/E_R=4$. Notice also the difference in time scales shown in the inset as the modes are found to thermalize around $t/T \sim 10$ in $Ng/E_R=8$ while thermalization is still absent in $Ng/E_R=4$ up to $t/T \sim 600$.

The apparent absence of thermalization in $Ng/E_R=4$ coincides with observations of incomplete level repulsion or nonchaoticity of the spectrum. Perhaps more importantly, it also correlates with the failure of the ETH due to the relative broadness of the EEV distribution around the initial energy [see upper left panel of Fig.~\ref{fig:eev}]. Later, we present an alternative picture in understanding  nonergodicity in the model by analyzing the dynamics of typical mean-field trajectories used in the ensemble averaging of the TWA.
We show in the upper panel of Fig.~\ref{fig:fockcohe2}, an example of decay of dynamical distance between a coherent and a Fock state with the same initial energies. The short-time oscillations in the dynamics of the mode population are akin to Bloch oscillations in the case of initial Bose-Einstein condensates loaded in tilted optical lattices \cite{Kolovsky2009,Kolovsky2010}. In the single-band limit, the Bloch oscillations are shown to decay exponentially in certain parameter regimes as a result of chaotic mean-field dynamics \cite{Kolovsky2009}. Here in the two-band limit, we observe similar exponential decay of these oscillations as the system approaches thermal equilibrium. As a matter of fact, similar to the single-band case, the mean-field trajectories are also chaotic whenever there is a decay in the oscillations of the mode occupations, which we discuss in Sec. \ref{sec:mfield}.

We now investigate the three-site case, $L=3$. We present typical relaxation of the mode occupation numbers for initial Fock states in Fig.~\ref{fig:focks3}. The time scale for the system to relax is now longer than that for $L=2$. Moreover, we find that the modes in the upper band relax at a faster (at least one order of magnitude) rate than the modes in the lower band. We also present in the lower panel of Fig.~\ref{fig:fockcohe2} the time evolution of $d^{\ell}_r(t)$ between initial Fock and coherent states for $L=3$, where it becomes more apparent that, indeed, the modes in upper band thermalizes much faster than those in the lower.

Figure \ref{fig:logdis3} shows the exponential decay of $d^{\ell}_r(t)$ for initial Fock states. This behavior is clearly seen in the slow decay of the dynamical distance of modes in the lower band. Furthermore, modes in the lower band are found to approach nonzero constant values of $d^{\ell}_r(t)$. The plateaus due to $d^{\ell}_r(t)$ being constant over an extended period of time can be interpreted as prethermalization plateaus over which the system becomes trapped in a metastable state. 
Similar to the two-site system, the $L=3$ case also do not fully thermalize for nonchaotic Hamiltonian such as $Ng/E_R=4$, which is depicted in the upper panel of Fig.~\ref{fig:logdis3}. Instead, the system gets trapped in extremely long-lived metastable state marked by small (relative to long-time limit) temporal fluctuations of $d^{\ell}_r(t)$ around some nonzero values over an extended time scale. During this period, memory of the initial conditions persists and therefore, thermalization is not yet fully attained. While for interactions that produce GOE level statistics as in $Ng/E_R=8$, the system thermalizes as the dynamical distance of all the modes in the system become zero in the long-time limit. 
Equilibration is also found in initial coherent states as shown in the lower panel of Fig.~\ref{fig:fockcohe2}.

\section{Properties of Mean-field trajectories}\label{sec:mfield}

In this section, we further study properties of the classical trajectories. In particular, we examine the dynamics at the level of a single mean-field trajectory corresponding to a solution of the Gross-Pitaevskii (GP) equation in Eq.~\eqref{eq:eom}. We then relate its qualitative behavior to the observed quantum dynamics from the preceding section. The connection between quantum dynamical behaviors and characteristics of the mean-field solution is naturally expected since individual trajectories in the TWA are solutions of the GP equation. 

\begin{figure}[!ht]
\begin{center}
\includegraphics[width=0.5\columnwidth]{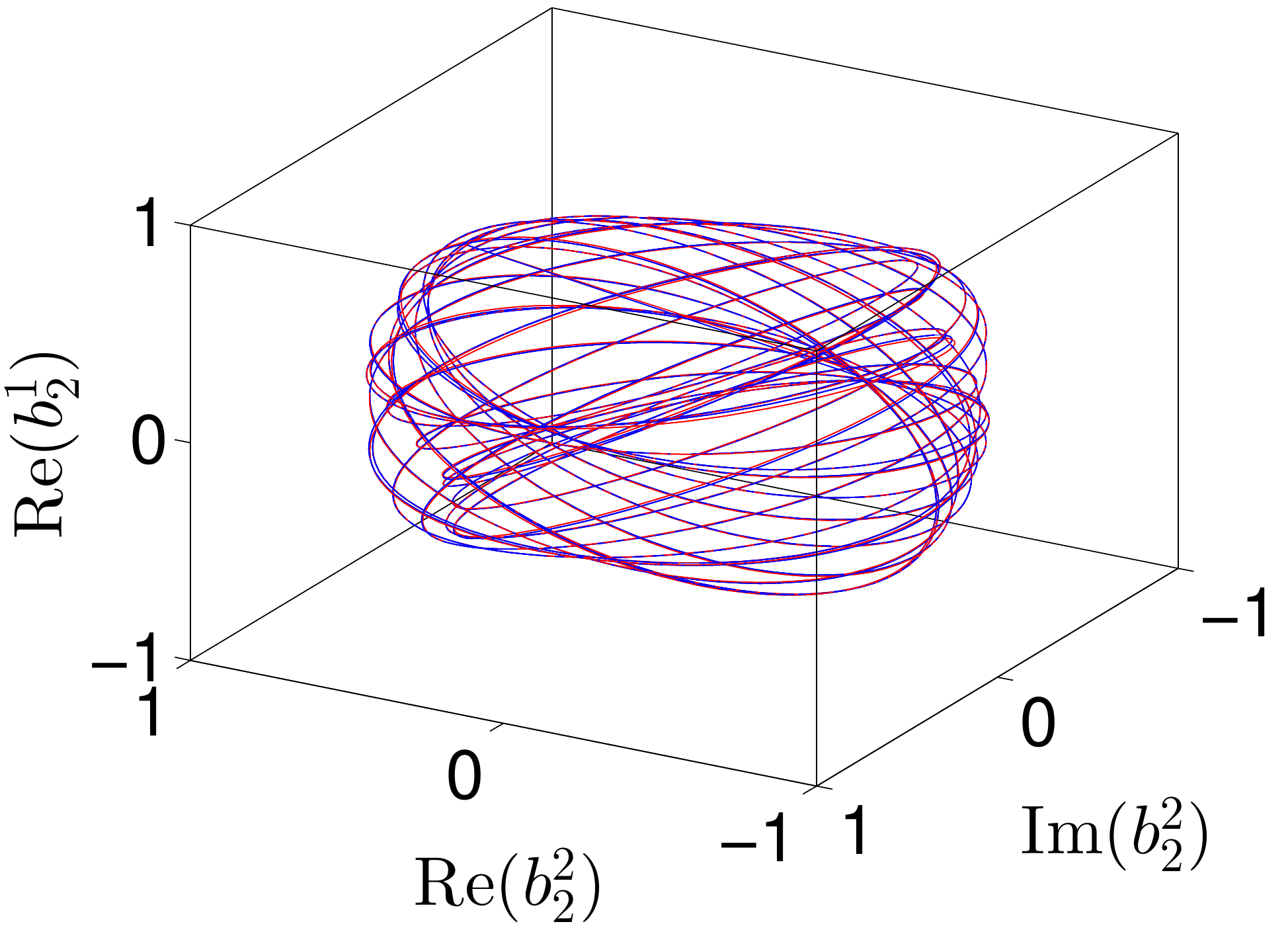}\includegraphics[width=0.5\columnwidth]{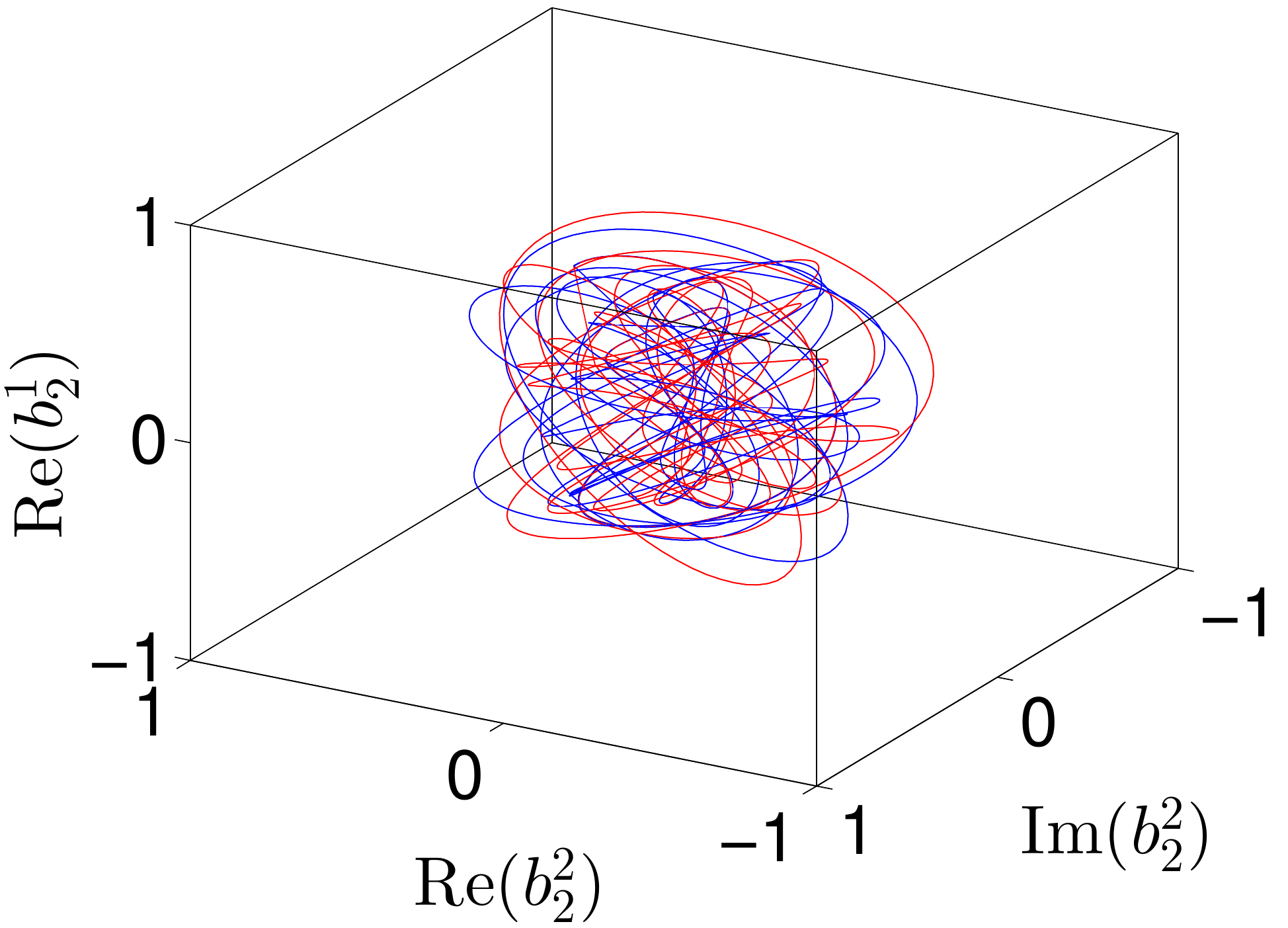}
\protect\caption{(Color online) Mean-field trajectories of two slightly different initial Fock states up to $t/T=20$ for $L=2$, (Left) $Ng/E_R=4$, and (Right) $Ng/E_R=8$. Note that we rescaled $b^{\ell}_r \to b^{\ell}_r/\sqrt{N}$.}
\label{fig:mffock2traj}
\end{center}
\end{figure}

\begin{figure}[!ht]
\begin{center}
\includegraphics[width=0.5\columnwidth]{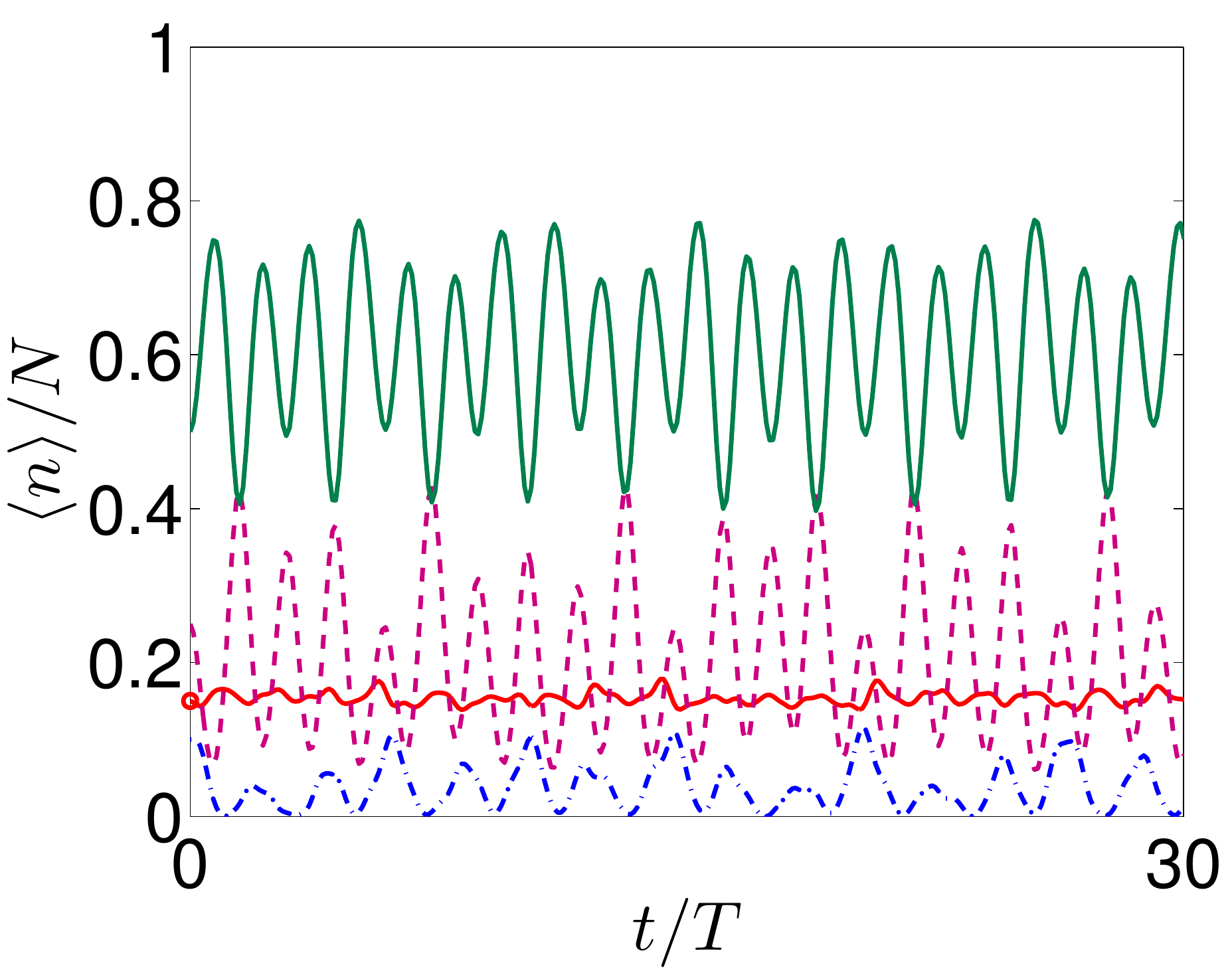}\includegraphics[width=0.5\columnwidth]{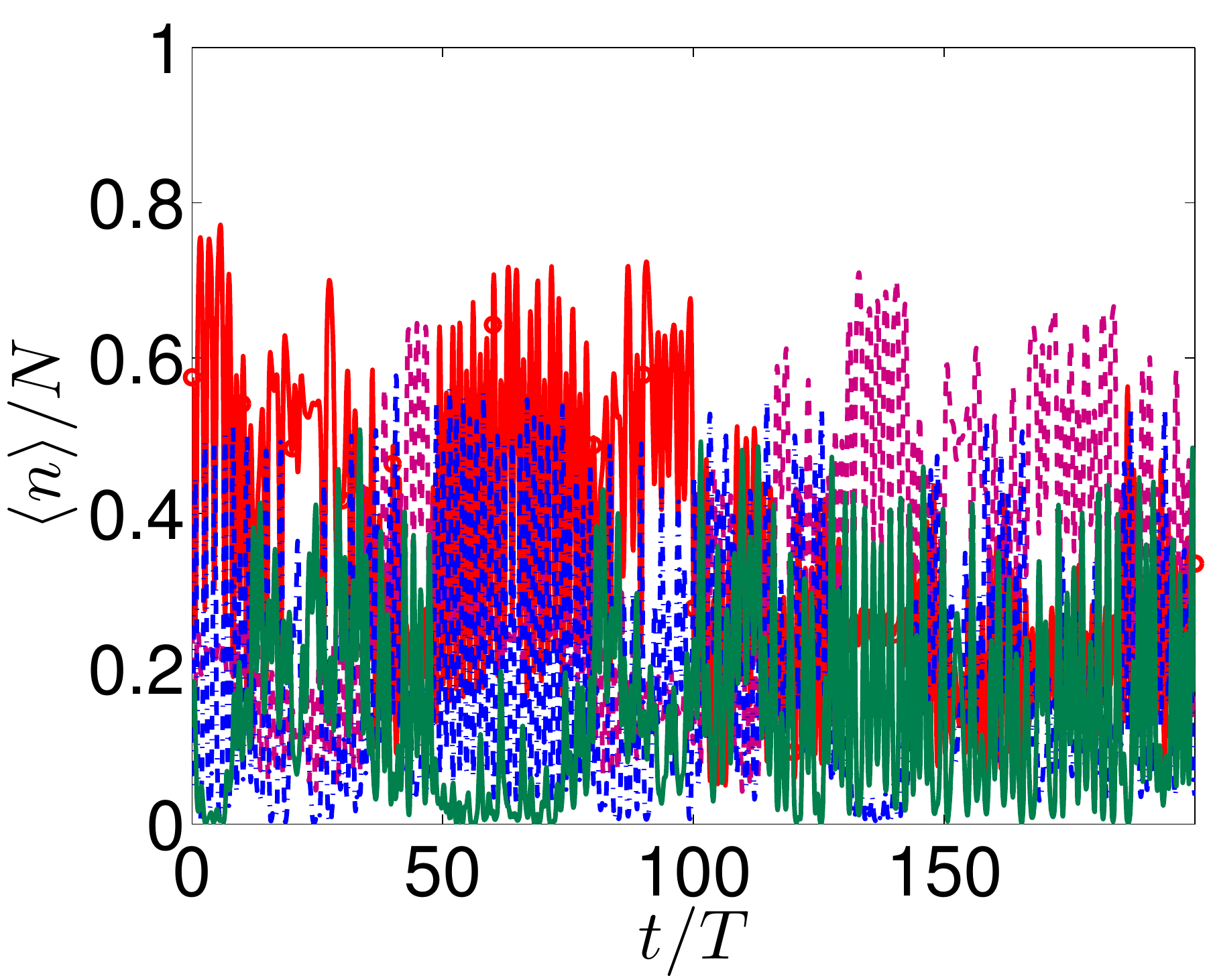}
\includegraphics[width=0.5\columnwidth]{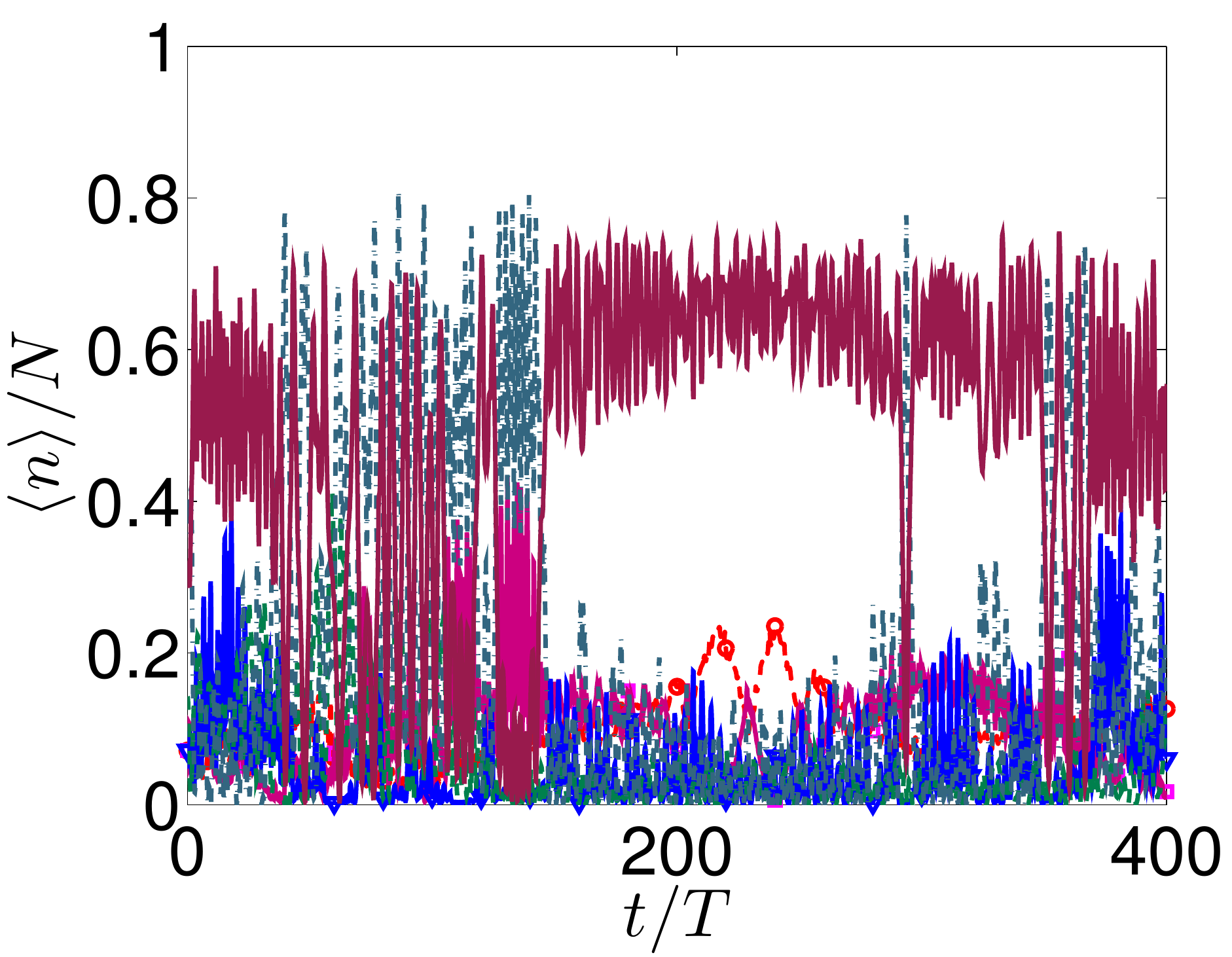}\includegraphics[width=0.5\columnwidth]{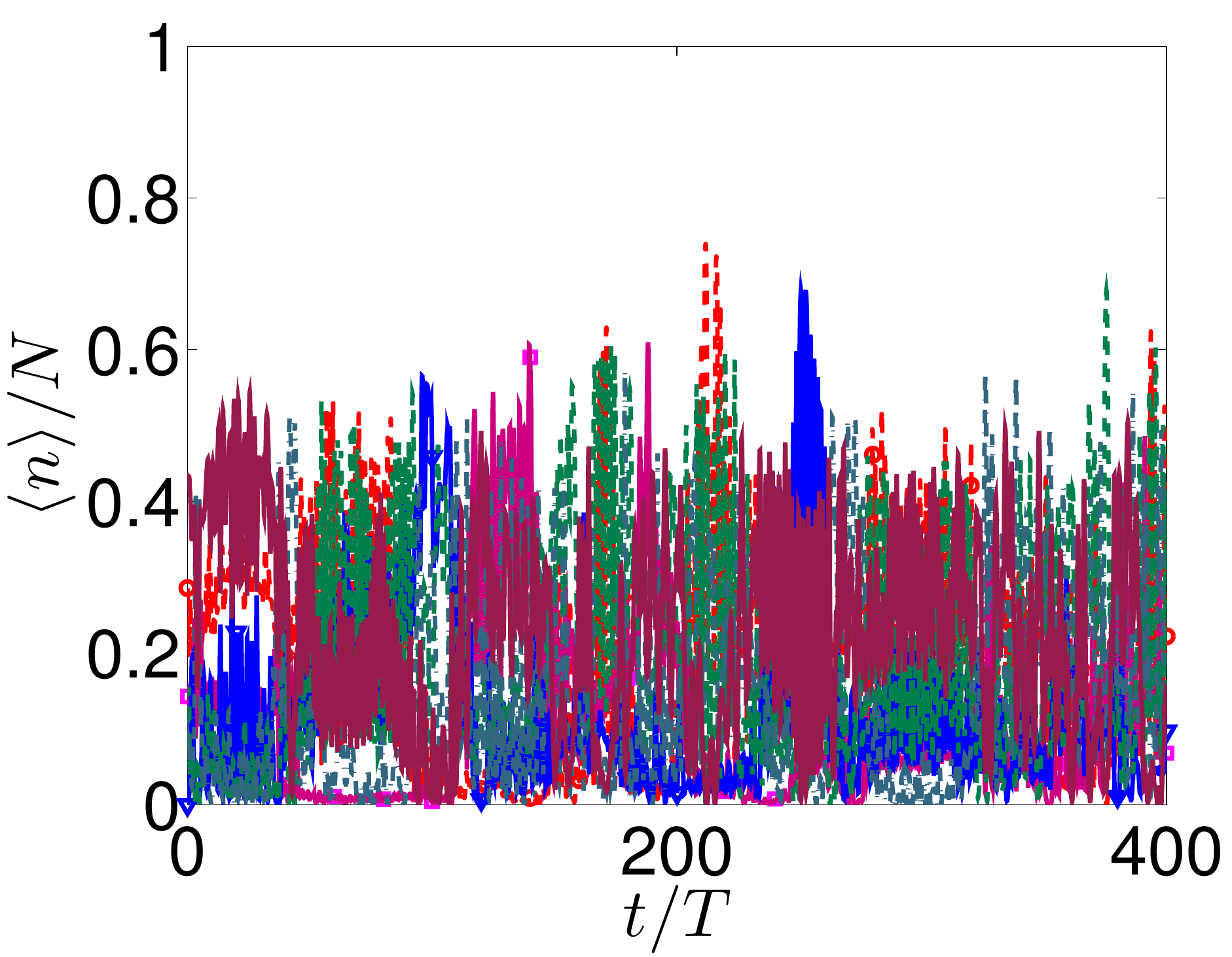}
\protect\caption{(Color online) Mean-field solution of Eq.~\eqref{eq:eom}, (Top) $L=2$; (Bottom) $L=3$; (Left column) $Ng/E_R=4$; (Right column) $Ng/E_R=8$. Individual curve in each plot represents the mean-field dynamics of a particular mode occupation number. Legend labels are the same as Fig.~\ref{fig:fockcohe2}.}
\label{fig:mffock2}
\end{center}
\end{figure}

We present in Fig.~\ref{fig:mffock2traj} the trajectories of two slightly different initial states. The initial states differ from each other by transferring 10 atoms from the lower mode of the second site to the lower mode of the first site. It can be seen that for $Ng/E_R=4$, the two trajectories do not diverge from each other as opposed to the chaotic nature of trajectories when $Ng/E_R=8$.
The chaoticity (regularity in the case of a nonergodic system) of trajectories is also evident from the mean-field dynamics of the mode occupation numbers. The mean-field dynamics for $L=2$ is shown in the top row of Fig.~\ref{fig:mffock2}. The upper-left panel of the same figure shows that the mean-field solution for $Ng/E_R=4$ can exhibit both self-trapping (lower mode of first site) and quasiperiodic oscillations (remaining three modes). The presence of both dynamical behaviors can also be found in mixed regions of classical phase space for the untilted multilevel double well system \cite{Gillet2014}. Nevertheless, an averaging over the ensemble of trajectories leads to the relaxation of mode occupation numbers to a nonthermal state as discussed before. In contrast, irregular or chaotic evolution of the mode populations is observed in quantum chaotic regime of strong interaction, for example $Ng/E_R=8$.

\begin{figure}[!ht]
\begin{center}
\includegraphics[width=0.75\columnwidth]{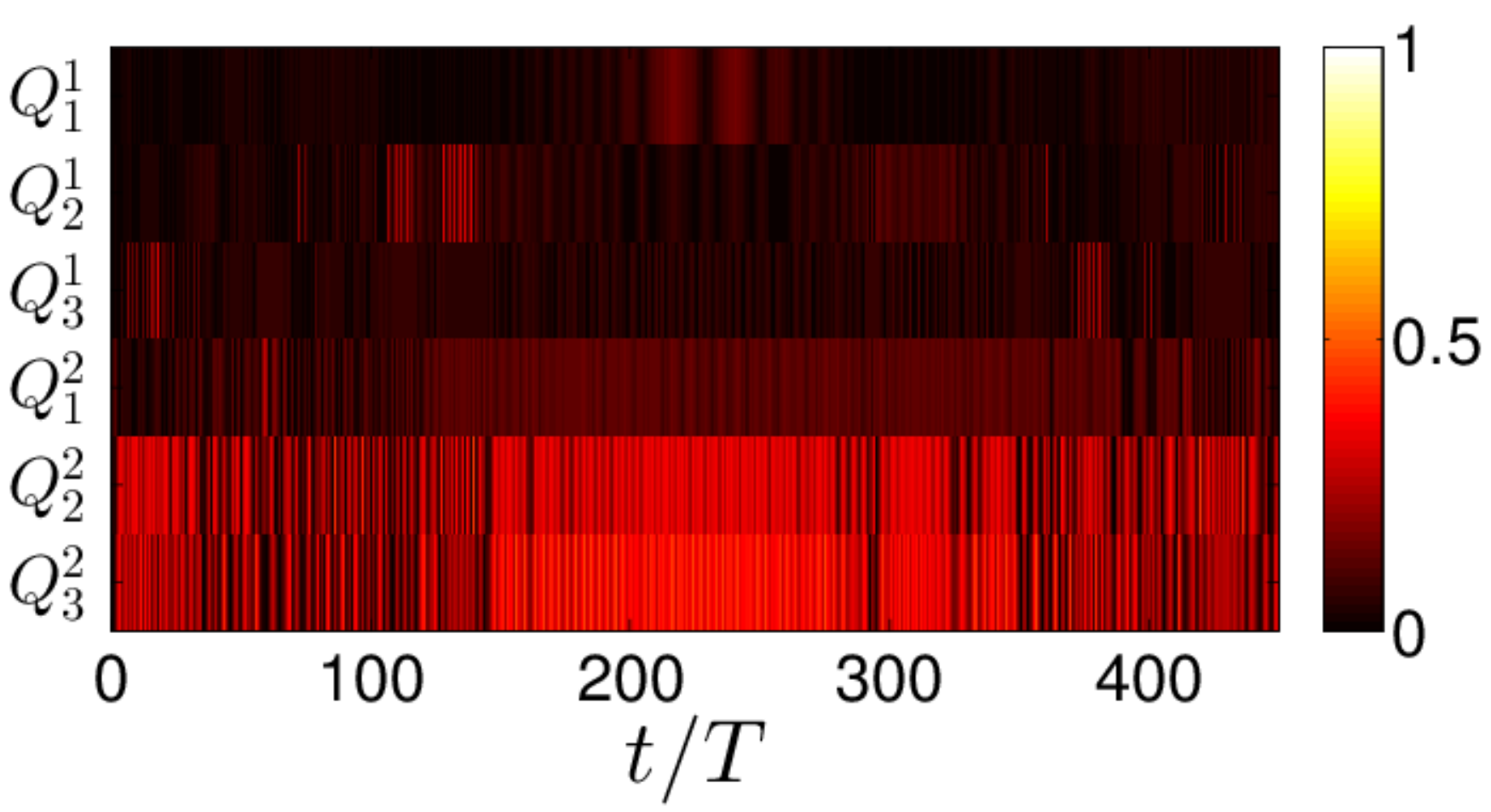}
\includegraphics[width=0.75\columnwidth]{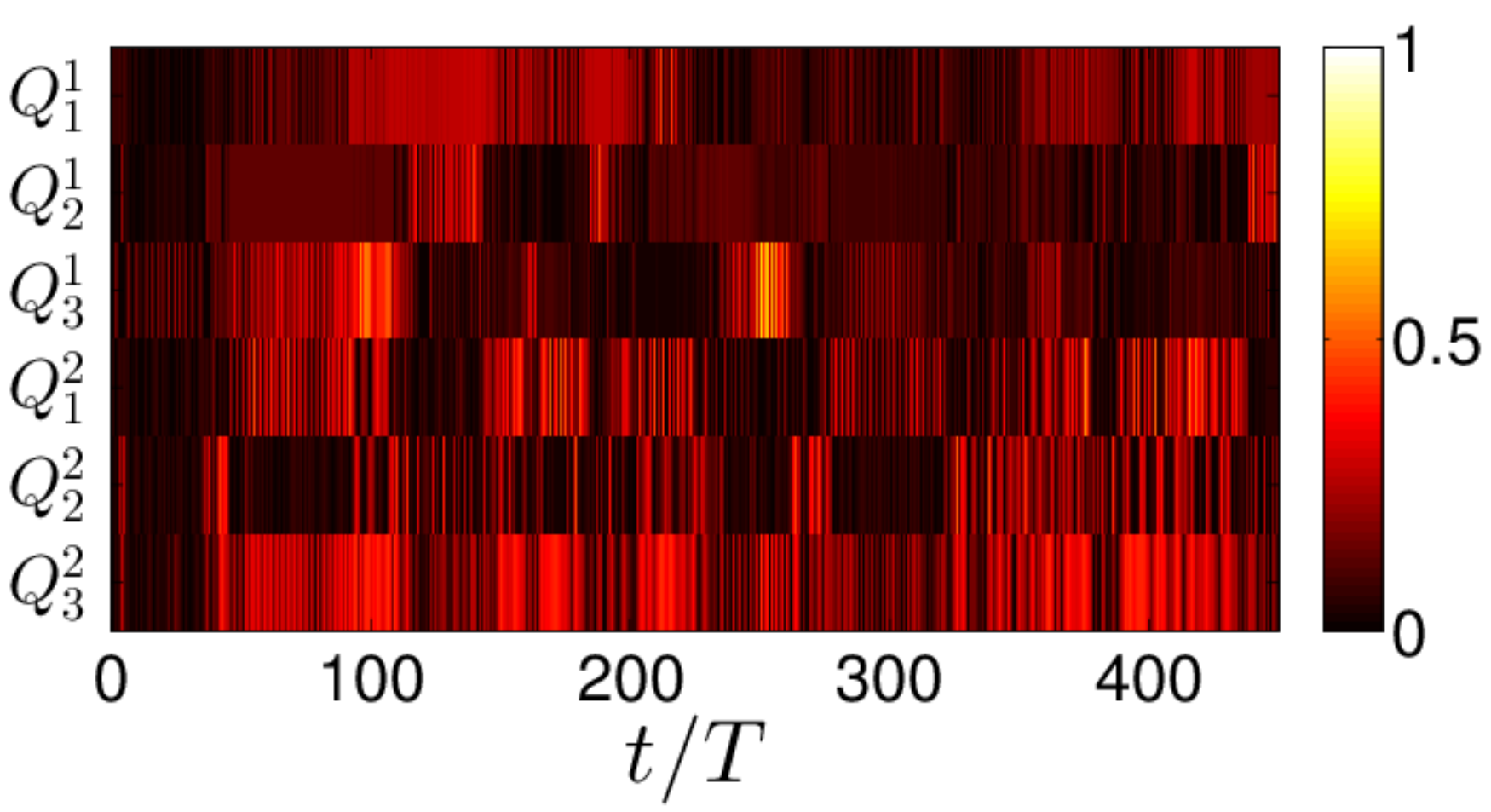}
\protect\caption{(Color online) Density plot of the local mobility in mode-time for representative mean-field trajectories with $L=3$. (Top) $Ng/E_R=4$ with Fock initial state $\{n^1_1,n^1_2,n^1_3,n^2_1,n^2_2,n^2_3\} = \{1,1,1,2,5,4\}\times 10^4 $; and (Bottom) $Ng/E_R=8$ with Fock initial state $\{n^1_1,n^1_2,n^1_3,n^2_1,n^2_2,n^2_3\} = \{2,1,5,1,3,2\}\times 10^4 $. }
\label{fig:mob}
\end{center}
\end{figure}

Recently, the emergence of metastable prethermalized states in the single-band Bose-Hubbard model has been studied in conjunction with the appearance of dynamical heterogeneity of the underlying classical trajectories in the TWA \cite{Landea2015}. The same perspective can be used in gaining further insights into the slow and hierarchical relaxation dynamics of the three-site system demonstrated in the preceding section.
The first step is to look at representative trajectories such as the ones shown in the bottom row of Fig.~\ref{fig:mffock2}. For $Ng/E_R=4$, it can be seen in the lower-left panel of Fig.~\ref{fig:mffock2} that the occupation numbers in the upper modes of the second and third sites are fluctuating around values larger than the initial population of these modes. On the other hand, the remaining modes are found to fluctuate closer to their initial population. 
We can then assign a measure of local mobility for each mode occupation numbers defined as
\begin{equation}
 Q^{\ell}_r(t) = | n^{\ell}_r(t) - n^{\ell}_r(0)  |/N,
\end{equation}
which quantifies the displacement of a classical trajectory with respect to its initial position \cite{Hedges2009,Chandler2010,Landea2015}. We illustrate in Fig.~\ref{fig:mob}, the density plot of the local mobility for representative trajectories when $L=3$. Dark (bright) areas in the density plot can be interpreted as regions with slow (fast) mobility.  Similar to the observations of Ref.~\cite{Landea2015}, features reminiscent of dynamical heterogeneity \cite{Hedges2009,Chandler2010} such as sharp contrast between regions of slow and fast mobility can also be found in the mode-time structure of a single trajectory to possess. This can be seen in both the upper panel (around $t/T \in [150,300]$) and the lower panel (time scales up until $t/T \sim 250$) of Fig.~\ref{fig:mob}. Due to the allowed coupling with modes in the upper band, dynamical heterogeneity in the system manifests in an extended configuration space in addition to the usual space-time degrees of freedom. 

Dynamical heterogeneity is one of the precursors to first-order dynamical phase transition in explaining the formation of classical glasses \cite{Hedges2009,Chandler2010}. Our next goal is to check if such dynamical phase transition can occur in the two-band system. If so, this can help us understand the multistep relaxation process observed in Fig.~\ref{fig:logdis3}, at least in the level of semiclassical dynamics. To this end, we proceed by calculating the activity of each classical trajectory $[b(t)]$ defined by
\begin{equation}\label{eq:kbt}
	K[b(t)] = N^{-4}(\Delta t/T) \sum_t^{t_{\mathrm{obs}}} \sum_{\ell=1}^2 \sum_{r=1}^L \biggl| |b^{\ell}_r(t+\Delta t)|^2 - |b^{\ell}_r(t)|^2 \biggr|^2,
\end{equation}
where $t_{\mathrm{obs}}$ is the maximum observation time of the mean-field trajectory \cite{Hedges2009,Chandler2010,Landea2015}. The summation over $t$ is done in incremental step, $\Delta t$, to disregard small oscillations in the $M+1$ mode-time configuration space. Note that in Eq.~\eqref{eq:kbt}, the $c$-numbers are rescaled by $b^{\ell}_r(t) \to b^{\ell}_r(t)/\sqrt{N}$. The activity $K[b(t)]$ is typically large for active or ergodic trajectories while it is typically small for inactive or nonergodic trajectories.

\begin{figure}[!ht]
\begin{center}
\includegraphics[width=1.0\columnwidth]{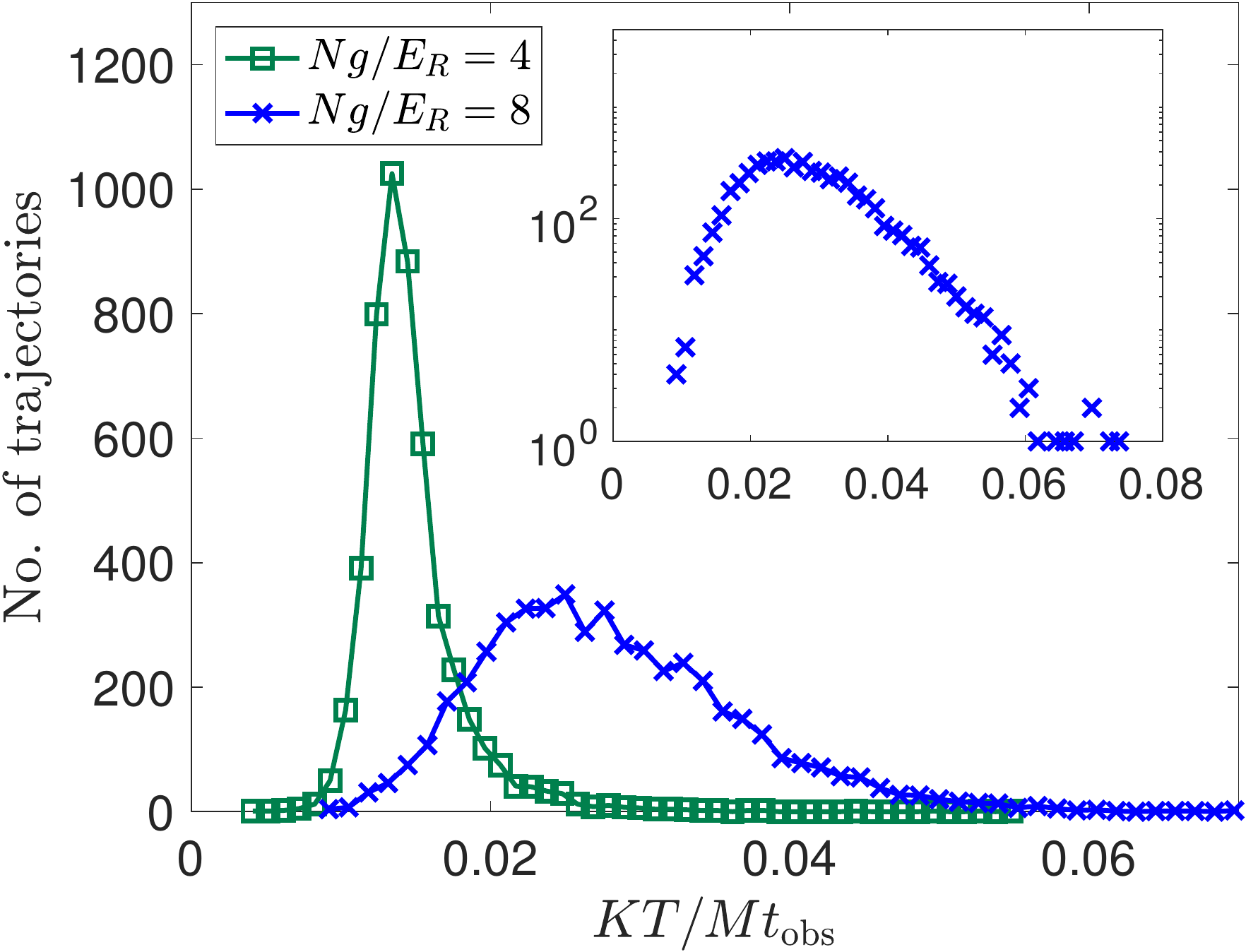}
\protect\caption{(Color online) Distribution of activity $K[\psi(t)]$ of the mean-field trajectories at $t_{\mathrm{obs}}/T=150$ for $L=3$ with the same initial states as in Fig.~\ref{fig:mob} and $\Delta t/T = 0.4$. Inset: Semilogarithmic scale of the distribution for $Ng/E_R = 8$.}
\label{fig:mobhist}
\end{center}
\end{figure}

\begin{figure}[!ht]
\begin{center}
\includegraphics[width=0.9\columnwidth]{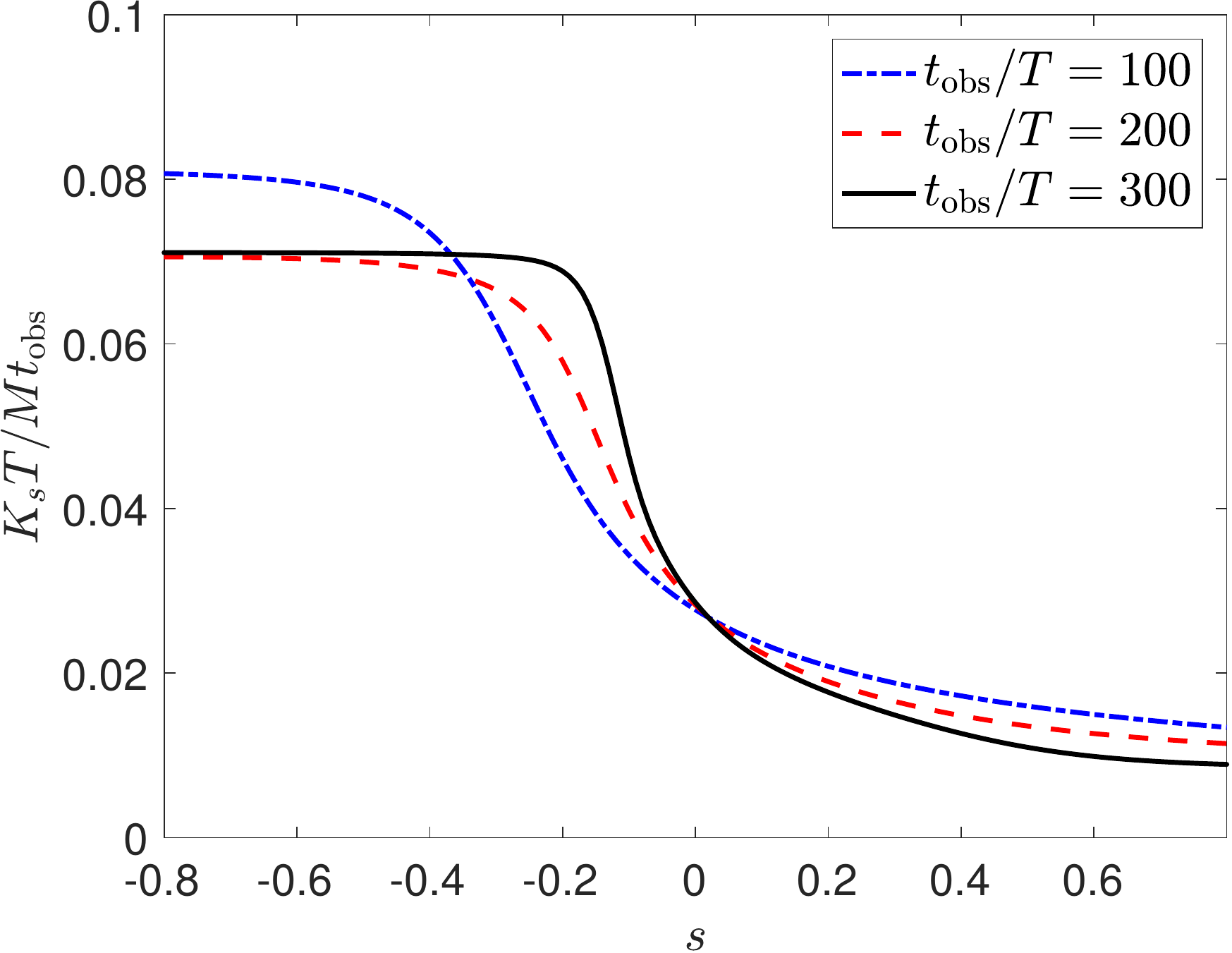}
\includegraphics[width=0.9\columnwidth]{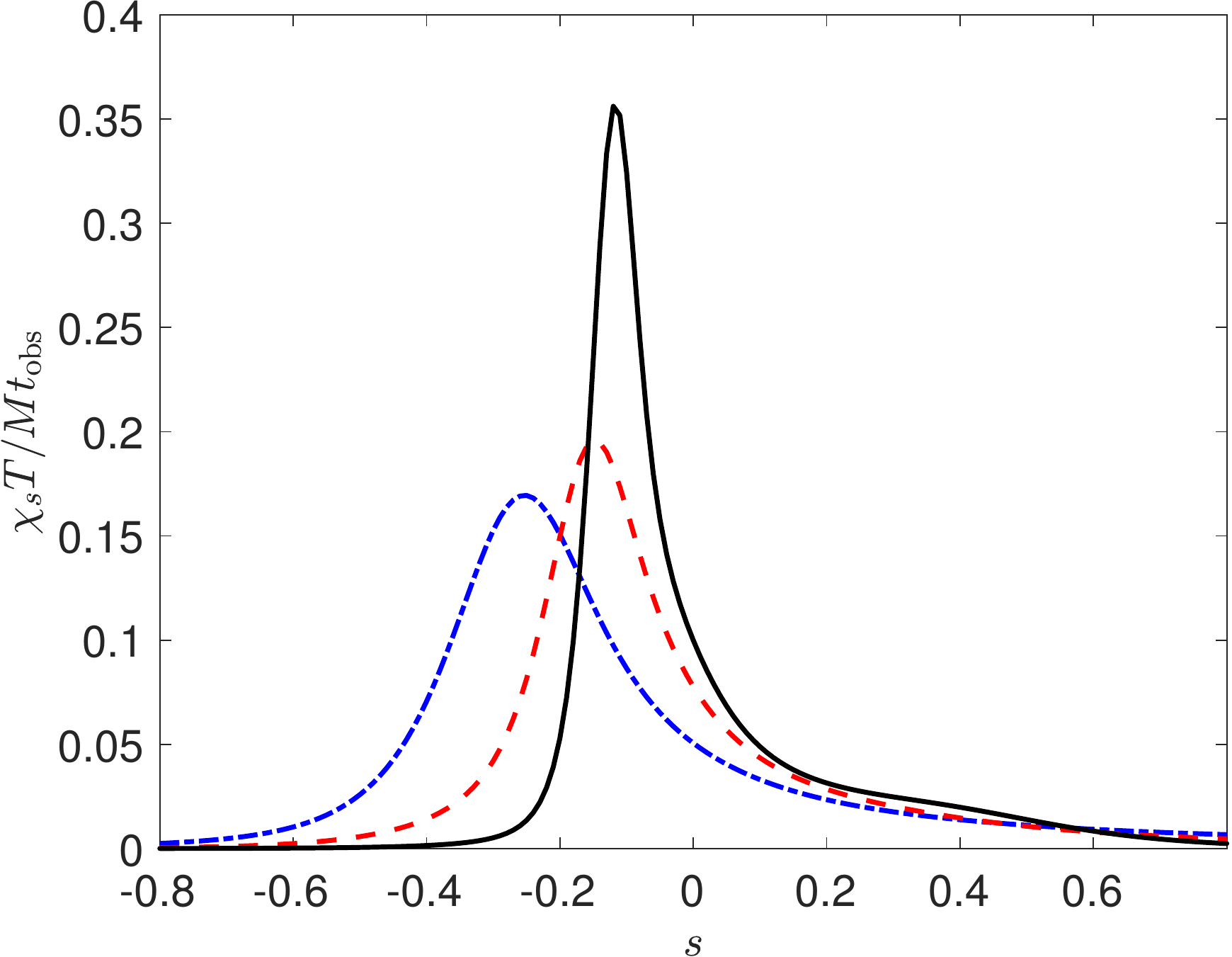}
\protect\caption{(Color online) (Top) Mean activity $K_s$ and (Bottom) susceptibility $\chi_s$ as a function of $s$ for different observation time $t_{\mathrm{obs}}$ for $L=3$, $Ng/E_R=8$, and the same initial Fock state as in Figs.~\ref{fig:mob} and \ref{fig:mobhist}.}
\label{fig:dpt}
\end{center}
\end{figure}

We illustrate in Fig.~\ref{fig:mobhist}, examples of the distribution of activity. The mean value of the activity, approximated by the position of the peak of the distribution, is clearly smaller for $Ng/E_R=4$ than $Ng/E_R=8$. Although in the case of $Ng/E_R=8$, a number of classical trajectories in the left-end tail of the distribution in Fig.~\ref{fig:mobhist} is found to have low activity.
This suggests the possibility of coexistence between nonergodic and ergodic trajectories with homogeneous structures in mode-time (see intermittent bright and dark regions in the lower panel of Fig.~\ref{fig:mob}). The set of ergodic trajectories drives the system towards a thermal state. The appearance of nonergodic trajectories in conjunction with ergodic ones can lead to hierarchical relaxation. 

We can further quantify the dynamical phase transition between ergodic and nonergodic trajectories. The relaxation dynamics will have to slow down as this transition is approached and this may cause the appearance of multistep relaxation.
The activity $K[b(t)]$ can act as an order parameter, which is then coupled to a field $s$ \cite{Garrahan2007,Hedges2009,Garrahan2010,Chandler2010,Landea2015}. The role of the field $s$ is analogous to the role of $\beta=1/k_{\mathrm{B}}T$ in statistical mechanics of equilibrium phase transitions \cite{Hedges2009}.
This procedure introduces a weight on the probability distribution of a trajectory $[b(t)]$ based on its activity,
\begin{equation}
	P_s[b(t)] = \frac{P_0[b(t)] e^{-sK[b(t)]}}{Z_s},
\end{equation}
where $Z_s$ is a partition function. We shall focus on initial Fock states, for which slow multistep dynamics is observed in Fig.~\ref{fig:logdis3}. We set the un-weighted distribution $P_0[b(t)]$ as the Wigner function of Fock states in the large $N$ limit. Note that we only consider initial product states and therefore, $P_0[b(t)]$ is just a product of Wigner functions in each mode
\begin{equation}
	W(b^\ell_r,b^{\ell *}_r) = \sqrt{\frac{2}{\pi}}\mathrm{exp}\biggl( -{2( |b^\ell_r|^2 - |b_0|^2 -1/2)^2} \biggr),
\end{equation} 
where $|b_0|^2$ is the occupation number of a mode \cite{Olsen2009}. The expectation value of the order parameter is obtained from
\begin{equation}
	K_s = \frac{1}{Z_s} \sum_{b(t)} P_s[b(t)] K[b(t)].
\end{equation}
The weight assigned on ergodic and nonergodic trajectories depends on the sign of $s$, i.e., for positive (negative) values of $s$, trajectories with high (low) activity are exponentially suppressed. Phase transition is observed when there is an abrupt change in the weighted average of the order parameter, $K_s$, as a function of $s$. When the observation time is increased, there is a marked crossover from ergodic to nonergodic behavior as the step in $K_s$ becomes steeper. This translates to an increase in the peak of the susceptibility,
\begin{equation}
	\chi_s = -\frac{\partial K_s}{\partial s} = \frac{1}{Z_s} \sum_{b(t)} P_s[b(t)] (K[b(t)]-K_s)^2.
\end{equation}

Recall that the dynamics for the set of initial Fock states, whose dynamics is shown in the lower panel of Fig.~\ref{fig:logdis3}, is characterized by a slow exponential decay of $d^{\ell}_r(t)$ towards a prethermal state and then followed by an eventual thermalization of the system. We demonstrate in Fig.~\ref{fig:dpt} a dynamical phase transition for this set of initial states, where the critical value  is found near $s \approx -0.1$. Note that in practice, the range of $s$ explored in Fig.~\ref{fig:dpt} has been chosen such that the most likely values of the distribution of activity shown in the inset of Fig.~\ref{fig:mobhist} are well inside the range of $K_s$ that have been sampled in the upper panel of Fig.~\ref{fig:dpt}.
Similar to the single-band Bose-Hubbard model in Ref.~\cite{Landea2015}, the phase transition in the two-band model appears to be continuous. The unimodal distribution of the activity in Fig.~\ref{fig:mobhist} is indicative of such continuous phase transition, in contrast with a bimodal distribution, which generically occurs for discontinuous phase transition both in classical \cite{Garrahan2007,Hedges2009} and quantum systems \cite{Garrahan2010}. Nonetheless, the peak close to $s=0$ of $\chi_s$ is a signature of the crossover between the active and inactive regimes where the ensemble of trajectories is a mixture of coexisting ergodic and nonergodic trajectories.

Finally, our findings suggest a possible mechanism for prethermalization in the context of classical trajectories for the two- and three-site models analyzed in this work which is the presence of inactive or nonergodic trajectories due to clustering of slow-fast dynamics of different modes, which mimics the space-time dynamical heterogeneity in theory of glasses. In the case of $L=2$, we have shown an early stage of this prethermalization mechanism where the dynamics is dominated by regular trajectories such as self-trapping and quasi-periodic oscillations. Nevertheless, it is clear for the case with $Ng/E_R = 4$ shown in the top panel of Fig. \ref{fig:focks2} that the dynamics is not purely regular and that there is already enough nonergodic trajectories coexisting with the regular ones to make the dynamics look irregular albeit not yet fully thermalized. On the other hand for $L=3$, we have seen the next stage of the prethermalization mechanism where nonergodic trajectories start to coexist with the ergodic ones.

\section{Conclusions}\label{sec:conc}

We have numerically studied relaxation processes in a tilted two-band generalization of the Bose-Hubbard model from the perspective of semiclassical dynamics. In particular, we have used the TWA in simulating the relaxation dynamics of the mode occupation numbers in the limit of large number of bosons and extremely large filling factor $N/L$ with $L = \{2,3\}$. By fixing all parameters except for the interaction strength $g$, we have demonstrated that the system do not properly thermalize if the quantum mechanical counterpart of the system does not possess a fully chaotic spectrum in the language of quantum chaos. While for chaotic spectrum, we have found initial state independence of the system through the decay of the dynamical distance between different initial Fock and coherent states with the same conserved quantities and, thus, thermalization is fulfilled in such cases. 
Evidence of metastable states in the system is demonstrated and analyzed from the perspective of the underlying mean-field trajectories in the TWA simulations. Furthermore, we have argued on the basis of statistical properties of the underlying classical trajectories that prethermalization in the system can be attributed to a phase transition between ergodic and nonergodic trajectories. This dynamical phase transition occurs in an extended configuration space, which now includes the modes in the upper band in addition to the space-time degrees of freedom. Our analyses of the properties of the mean-field dynamics suggest that in the semiclassical limit of large filling factors, initial state independence within single-particle observables can be shown to dynamically emerge due to the uncertainty introduced by quantum fluctuations. Chaoticity of each mean-field trajectory is a necessary but not sufficient condition for dynamical relaxation to occur. However, an ensemble averaging of highly mobile and chaotic trajectories representing the initial quantum phase-space distribution will allow for manifestation of prethermalization or thermalization.

\acknowledgments

This work was supported by Massey University, by means of the Massey University Doctoral Research Dissemination Grant.

\bibliography{ref}

\end{document}